\documentclass[12pt]{article}

\usepackage{latexsym,amsmath,amsfonts,amssymb}
\usepackage{graphicx}
\usepackage{bbm}
\usepackage{cite}
\newcommand{\im}{\,\mathbb{I}\mbox{m}\,}

\newcommand{\Nugual}[1]{$\mathcal{N}\!= #1 $}

\newcommand{\eg}{\textit{e.g.}}

\newcommand{\ie}{\textit{i.e.}}

\newcommand{\rep}[1]{\mathbf{#1}}
\newcommand{\mr}{\mathrm}

\numberwithin{equation}{section}

\newcommand{\be}{\begin{equation}} \newcommand{\ee}{\end{equation}}
\newcommand{\bea}{\begin{equation} \begin{aligned}} \newcommand{\eea}{\end{aligned} \end{equation}}

\newcommand{\calM}{\mathcal{M}}

\newcommand{\bbC}{\mathbb{C}}
\newcommand{\bbH}{\mathbb{H}}

\newcommand{\bbP}{\mathbb{P}}
\newcommand{\bbR}{\mathbb{R}}
\newcommand{\bbZ}{\mathbb{Z}}

\immediate\write18{open -g -a spires \jobname.tex}
\def\inc#1#2{{%
\setbox1=\hbox{\includegraphics[scale=#1]{#2}}%
\raise-.5\ht1\box1%
}}

\def\SU{SU}
\def\U{U}
\def\USp{USp}
\def\cN{\mathcal{N}}

\def\bC{\mathbb{C}}
\def\bZ{\mathbb{Z}}

\begin{document}

\begin{titlepage}

\rightline{PUPT-2303}
\vspace{0.8cm}
\begin{center}
{\LARGE{\bf Webs of Five-Branes and \\[5mm]
$\cN=2$ Superconformal Field Theories\\[20mm]}}

{{\large Francesco Benini, \qquad Sergio Benvenuti}  \\[5mm]
\textit{Department of Physics, Princeton University \\
\vspace*{-2pt} Princeton, NJ 08544, USA}

\bigskip
\medskip
\medskip
\medskip
\medskip

{\large and Yuji Tachikawa} \\[5mm]
\textit{School of Natural Sciences, Institute for Advanced Study \\
\vspace*{-2pt} Princeton, NJ 08540, USA}
}

\medskip

\medskip

\medskip

\medskip

\medskip

\vskip 20pt
\begin{minipage}[h]{16.0cm}
We describe configurations of 5-branes and 7-branes
which realize, when compactified on a circle,
new isolated four-dimensional $\cN=2$ superconformal field theories
recently constructed by Gaiotto.
Our diagrammatic method allows
to easily count the dimensions of Coulomb and Higgs branches,
with the help of a generalized s-rule.
We furthermore show that superconformal field theories with $E_{6,7,8}$ flavor symmetry
can be analyzed in a uniform manner in this framework; in particular we realize these theories at infinitely strongly-coupled limits of quiver theories with $SU$ gauge groups.

\end{minipage}
\end{center}

\end{titlepage}

{\small \tableofcontents}

\section{Introduction}

Brane constructions in string or M-theory
can tell us a great deal of non-perturbative information
about supersymmetric gauge theories. For example,
four-dimensional $\cN=2$ supersymmetric quiver gauge theories
can be implemented using a system of D4-branes suspended between
NS5-branes in type IIA string theory.
This configuration can be lifted to M-theory, in which D4- and NS5-branes
merge into a single M5-brane, physically realizing the Seiberg-Witten curve
which governs the low energy dynamics of the system \cite{Witten:1997sc}.

\begin{figure}
\centerline{
\hspace{\stretch{1}}
\includegraphics[scale=.5]{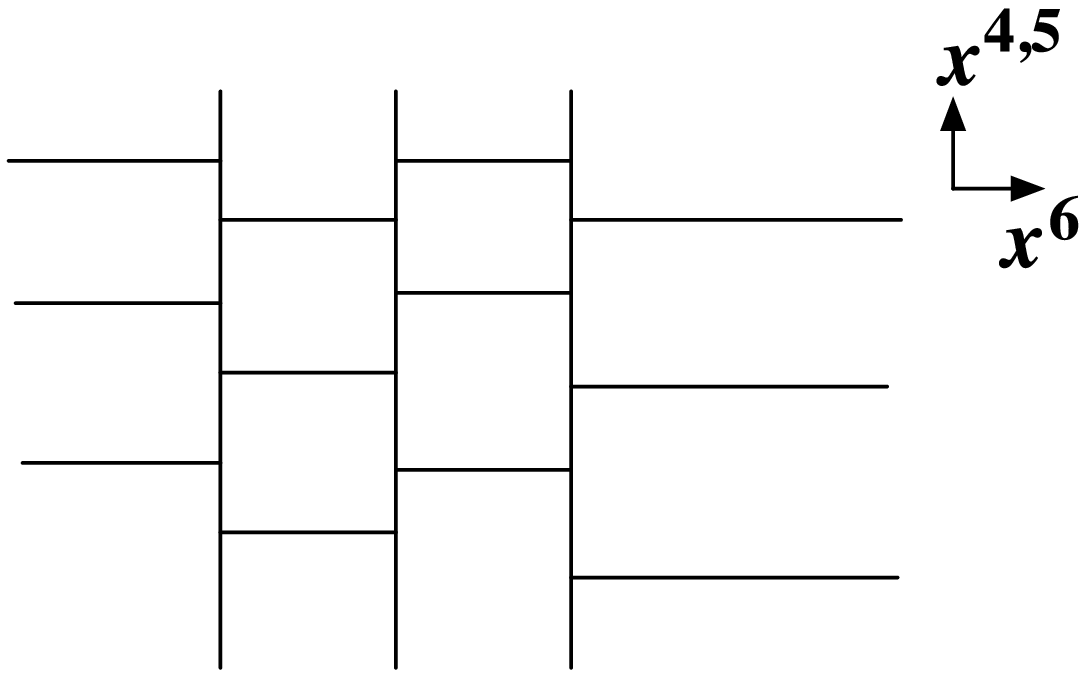}
\hspace{\stretch{2}}
\includegraphics[scale=.5]{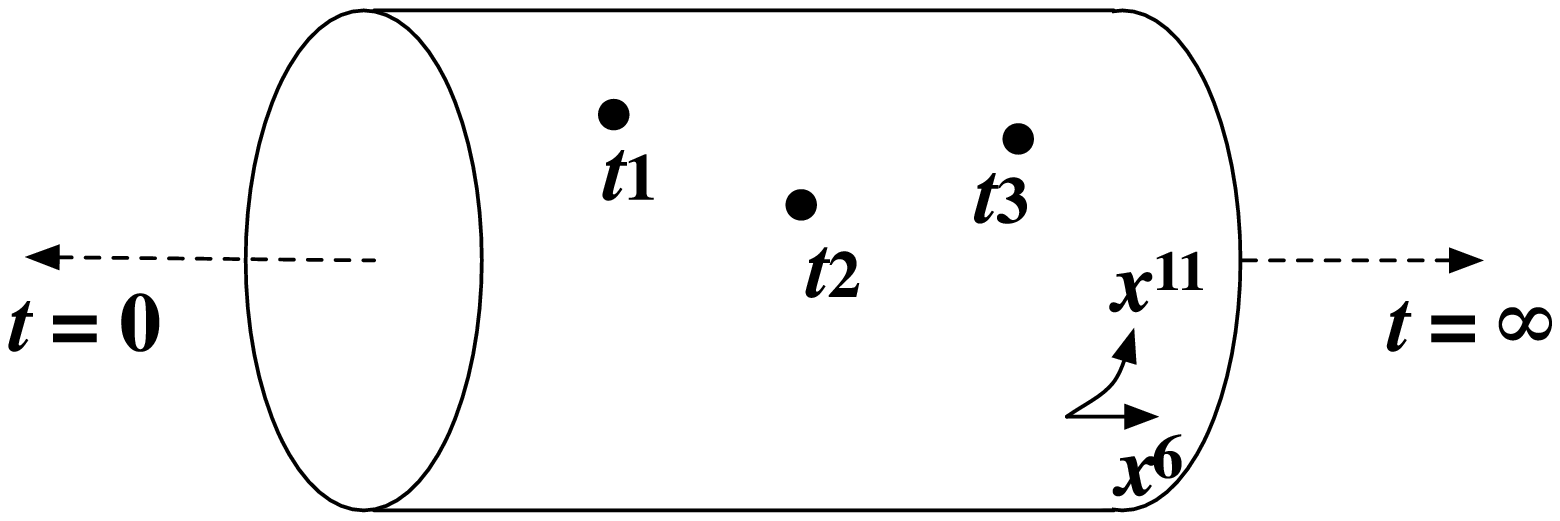}
\hspace{\stretch{1}}
}
\caption{\small Left: a brane configuration in type IIA. Vertical lines are NS5-branes
and horizontal lines are D4-branes.
Right:
its lift to M-theory showing the M-theory circle. \label{tac:IIA}}
\end{figure}

The setup is schematically drawn in Fig.~\ref{tac:IIA}.
Vertical lines stand for NS5-branes extending along $x^{4,5}$ while
horizontal lines are D4-branes extending along $x^6$, suspended between two NS5-branes,
or ending on an NS5-brane and extending to infinity.
All branes fill the space-time, $x^{0,1,2,3}$.
The example shown has two $\SU(3)$ gauge groups, each with three fundamental hypermultiplets, and one bifundamental hypermultiplet charged under the two gauge groups.

The system lifts to a configuration of M5-branes in M-theory.
It is natural to combine the direction along the M-theory circle, $x^{11}$, with
the direction $x^6$ to define a complex coordinate $t=\exp(x^6+i x^{11})$.
In this particular example, when all of the vacuum expectation values (VEV's) of the adjoint scalar fields
are zero, three M5-branes wrap the cylinder parameterized by $t$,
and at three values of $t$, say $t=t_{1,2,3}$, the stack of three M5-branes
is intersected by one M5-brane.

\begin{figure}
\centerline{
\hspace{\stretch{1}}
\includegraphics[scale=.4]{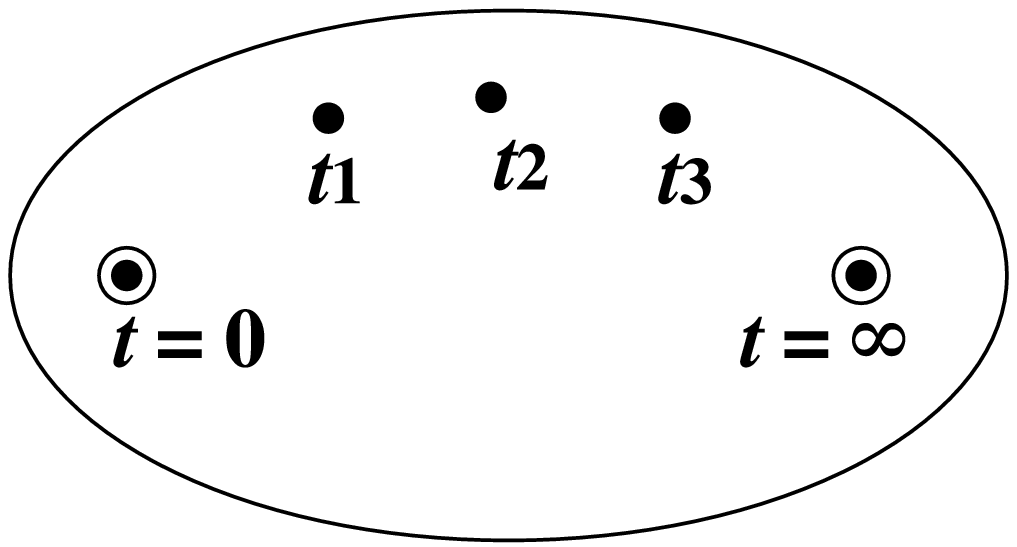}
\hspace{\stretch{2}}
\includegraphics[scale=.4]{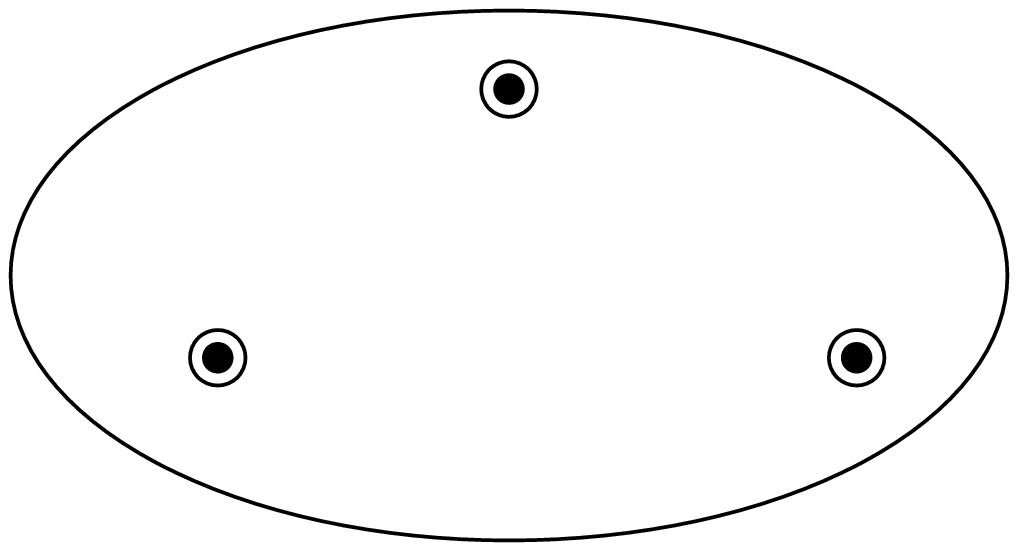}
\hspace{\stretch{1}}
}
\caption{\small Left: the system in Fig.~\ref{tac:IIA} as the compactification of M5-branes on a sphere with defects. The symbol $\bullet$ marks the {\em simple punctures}, and $\odot$  the {\em full punctures}.  Right: the compactification of M5-branes corresponding to the $T[A_{N-1}]$ theory, with three full punctures.
It has no obvious type IIA realization. \label{tac:M}}
\end{figure}

It was recently shown in \cite{Gaiotto:2009we} that
the system can also be seen as a compactification of $N$ M5-branes on a sphere
by a further change of coordinates which is only possible when
all of the gauge couplings are marginal. The resulting
configuration is shown on the left of  Fig.~\ref{tac:M}.
In this representation, both the intersections with other M5-branes, and the two infinite ends
can be thought of as conformal defect operators on the worldvolume of the M5-branes.
We call the defect corresponding to the intersection with another M5-brane the {\em simple puncture},
and the defect corresponding to intersecting $N$ semi-infinite M5-branes the {\em full} or {\em maximal puncture}.
It was found that marginal coupling constants are encoded by the positions of
the punctures on the sphere.

From this point of view, one can  consider
compactifications of $N$ M5-branes with more general configurations of punctures.
The most fundamental one is the sphere with  three full punctures,
depicted on the right hand side of Fig.~\ref{tac:M}. This theory, called $T[A_{N-1}]$,
is isolated in that it has no marginal coupling constants
because three points on a sphere do not have moduli.
It has (at least) $\SU(N)^3$ flavor symmetry, because each full puncture carries
an $\SU(N)$ flavor symmetry, as shown in \cite{Gaiotto:2009we}.
It arises in an infinitely strongly-coupled limit of a linear quiver gauge theory,
as a natural generalization of Argyres-Seiberg duality \cite{Argyres:2007cn}.
Furthermore, it is the natural building block from which
the four-dimensional superconformal field theory corresponding to
the compactification of $N$ M5-branes on a higher-genus Riemann surface can be
constructed as a generalized quiver gauge theory.
Its gravity dual  was found in \cite{Gaiotto:2009gz}.

It is clearly important to study the properties of this theory further, and
it will be nicer to have another description of the same theory
from which the different properties can be understood easily.
One problem is that this theory no longer
has a realization as a brane configuration in type IIA  string theory.
It is basically because the direction $x^6$ has only two ends, which can account
at most two special punctures on the sphere.
Instead, we propose that configurations of intersecting D5-, NS5- and (1,1) 5-branes
in type IIB string theory give the five-dimensional version of $T[A_{N-1}]$,
in the sense that compactification on $S^1$ realizes the theory $T[A_{N-1}]$.
We will see that each of the punctures corresponds to
a bunch of $N$ D5-branes, of $N$ NS5-branes, or of $N$ (1,1) 5-branes.

The realization of the field theory through a web of 5-branes makes manifest the moduli space, as happens with the more familiar type IIA construction \cite{Elitzur:1997fh, Hanany:1996ie}.
The Coulomb branch corresponds to normalizable deformations of the web which do not change the shape at infinity.
The Higgs branch  can be seen by terminating
all semi-infinite 5-branes on suitable 7-branes:
it then corresponds to moving the endpoints of 5-branes around,
as was the case in type IIA with D4-branes ending on D6-branes.

It was shown in \cite{Gaiotto:2009gz} that there are more general punctures or defects one can insert on the M5-brane worldvolume,
naturally labeled by Young tableaux consisting of $N$ boxes.
This kind of classification arises straightforwardly from the web construction, once 7-branes have been introduced:
we can group $N$ parallel 5-branes into smaller bunches, composed by $k_i$ 5-branes,
and then end $k_i$ 5-branes on the $i$-th 7-brane.
This leads to a classification in terms of partitions of $N$, in fact labeled by Young tableaux with $N$ boxes.
Recall that D5-branes terminate on D7-branes,
NS5-branes on $[0,1]$ 7-branes, and $(1,1)$ 5-branes on $[1,1]$ 7-branes.
Therefore the resulting system has mutually non-local 7-branes,
which are known to lead to enhanced symmetry groups when their combination
is appropriate \cite{DeWolfe:1998eu}.
This last observation will naturally lead us to propose
5-brane configurations which realize five-dimensional theories
with $E_{6,7,8}$ flavor symmetry, originally discussed in \cite{DeWolfe:1999hj}.
Our proposal is, as stated above, that these configurations are
five-dimensional versions of the theories in \cite{Gaiotto:2009we}.
It then gives a uniform realization of the four-dimensional superconformal field theories
with $E_{6,7,8}$ symmetry \cite{Minahan:1996fg,Minahan:1996cj} in the framework of \cite{Gaiotto:2009we}.
In particular it provides a new realization of the $E_{7,8}$ theories
using a quiver gauge theory consisting only of $\SU$ groups,
along the line of Argyres, Seiberg and Wittig \cite{Argyres:2007cn,Argyres:2007tq}.
In order to study configurations with general punctures,
application of the s-rule \cite{Hanany:1996ie} will be crucial.
We will need a generalized version of the s-rule studied in \cite{Mikhailov:1998bx,DeWolfe:1998bi,Bergman:1998ej} in the context of string junctions,
which we will review in detail, and its ``propagation'' inside the 5-brane web.

Finally, the web construction makes it clear that a theory with generic punctures
can arise as an effective theory by moving along the  Higgs branch of the $T[A_{N-1}]$ theory.
When two 7-branes are aligned in such a way that the 5-branes ending on them overlap, there can be Higgs branch directions corresponding to breaking the 5-branes on the 7-branes and moving the extra pieces apart. By moving the extra pieces very far away, one is left with a puncture with multiple 5-branes ending on the same 7-branes,
thus realizing generic punctures.

The paper is organized as follows:
we start in Sec.~\ref{sec: TN} by considering a junction of
$N$ D5-, NS5-, and $(1,1)$ 5-branes,
which we argue is the five-dimensional version of $T[A_{N-1}]$.
We study the flavor symmetry and the dimensions of Coulomb and Higgs branches.
To see the Higgs branch, we need to terminate the external
5-branes on appropriate 7-branes.
We proceed then in Sec.~\ref{sec: punctures}
to study how we can use 7-branes to terminate 5-brane junctions,
realizing more general type of punctures.
The s-rule governing the supersymmetric configurations of these systems
will also be formulated in terms of a dot diagram, that we will describe.
Several examples illustrating the generalized s-rule will be detailed in Sec.~\ref{sec: examples},
which naturally leads to our identification of certain 5-brane configurations
as the five-dimensional theories with $E_{6,7,8}$ flavor symmetry.
In Sec.~\ref{sec: davide}, which might be read separately,
we provide further checks of this identification
using the machinery in \cite{Gaiotto:2009we}, by showing that
the SCFTs with $E_{6,7,8}$ flavor symmetry arise in the strongly-coupled
limit of quiver gauge theories with $SU$ gauge groups.
We conclude with a short discussion in Sec.~\ref{sec: directions}. In App.~\ref{sec: SW curve}
we write down the Seiberg-Witten curve for the theory on the multi-junction.
%In App.~\ref{sec: generalized-s-rule} we review the derivation of the
%generalized s-rule.
Finally in
App.~\ref{sec: exp theories} we review some aspects of the $E_n$ theories.

% ---------------------------------------------------------------------------------
% ---------------------------------------------------------------------------------
% ---------------------------------------------------------------------------------

\section{$N$-junction and $T[A_{N-1}]$}\label{sec: TN}
\subsection{$N$-junction}
\begin{figure}[tn]
\centerline{
\includegraphics[height=17ex]{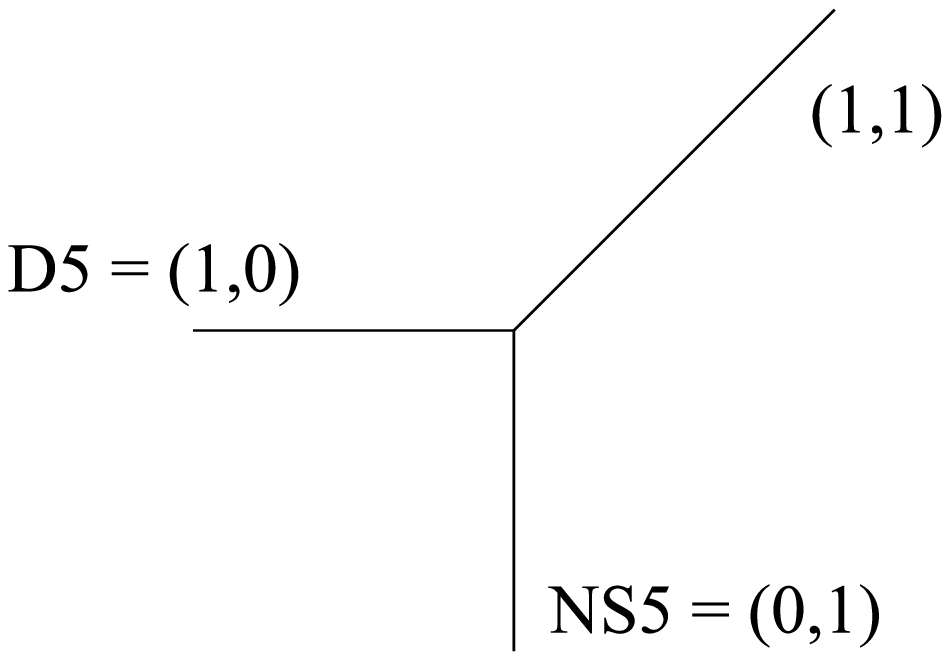}
\hspace{\stretch{1}}
\includegraphics[height=18ex]{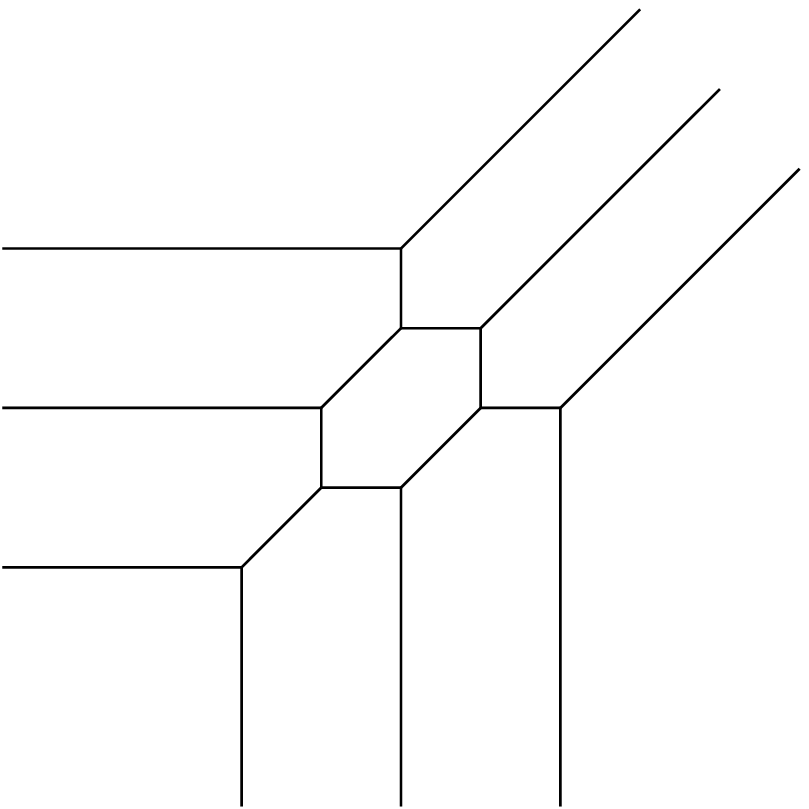}
\hspace{\stretch{1}}
\includegraphics[height=17ex]{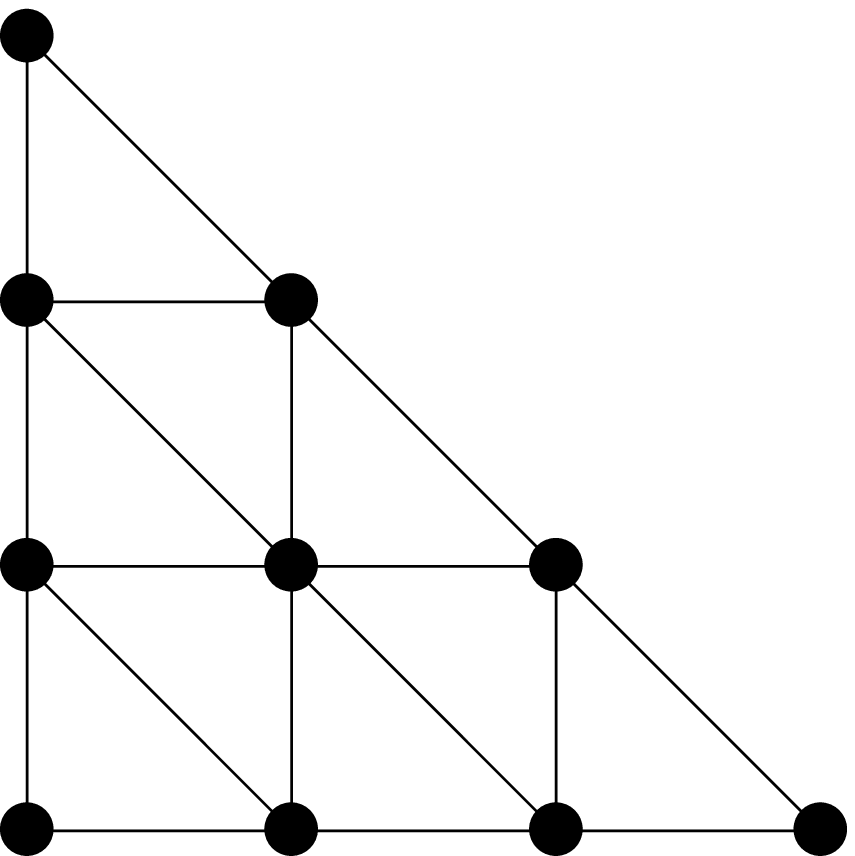}
}
\caption{\small Left: single junction of a D5, an NS5 and a $(1,1)$ 5-brane. Center: multi-junction of three bunches of $N=3$ 5-branes -- this realizes the $E_6$ theory. Right: the dual toric diagram of $\bbC^3/\bbZ_N \times \bbZ_N$, with a particular triangulation.
\label{fig: junctions}}
\end{figure}

\begin{table}[tn]
\centerline{
\begin{tabular}{c|cccc|c|cc|ccc}
& 0 & 1 & 2 & 3 & 4 & 5 & 6 & 7 & 8 & 9 \\
\hline
D5 & $-$ & $-$ & $-$ & $-$ & $-$ & $-$ & \\
NS5 & $-$ & $-$ & $-$ & $-$ & $-$ & & $-$ & \\
$(1,1)$ 5-brane & $-$ & $-$ & $-$ & $-$ & $-$ & \multicolumn{2}{c|}{angle} \\
7-branes & $-$ & $-$ & $-$ & $-$ & $-$ & & & $-$ & $-$ & $-$
\end{tabular}
}
\caption{\small Configuration of suspended $(p,q)$ 5-brane webs. To get a 4d theory, the direction $x^4$ is compactified on a circle. The symbol $-$ signifies that the brane extends in the corresponding direction.\label{tab: web config}}
\end{table}
We begin by summarizing the type IIB or F-theory configuration we will use in
Table~\ref{tab: web config}. There, the symbol $-$ under the column
labeled by a number $i$ means that the brane extends along the direction $x^i$.
The most basic object in the brane-web construction
is the junction between a D5-brane, an NS5-brane and a $(1,1)$ 5-brane \cite{Aharony:1997ju},
see Figure~\ref{fig: junctions}.
This system is rigid and does not allow any deformation, apart from the center of mass motion.
Accordingly, it does not give rise to any 5d low energy dynamics apart from the decoupled center of mass.

We can consider a configuration, which we refer to as the $N$-junction,
where $N$ D5-branes, $N$ NS5-branes and $N$ $(1,1)$ 5-branes meet,
see the diagram in the center of Figure~\ref{fig: junctions}.
For a given 5-brane web, the dual diagram is formed by
associating one vertex to each face, both compact and non-compact,
in the original brane web,
and connecting two vertices whenever the corresponding faces in the original
diagram are adjacent, see the rightmost diagram of Figure~\ref{fig: junctions}.
It is known that this procedure produces the toric diagram
of a non-compact Calabi-Yau threefold, and that M-theory
compactified on this threefold is dual to the original five-brane construction.
Under this duality, the single junction corresponds to the flat space $\bC^3$
whereas
the multi-junction of $N$ D5-, NS5- and (1,1) 5-branes
corresponds to the blow-up of the orbifold $\bbC^3/\bbZ_N \times \bbZ_N$
where $\bbZ_N\times \bbZ_N$ acts on $(x,y,z)\in \bbC^3$ by \begin{equation}
(x,y,z)\to (\alpha x,\beta y,\gamma z).
\end{equation}  Here $\alpha,\beta,\gamma$ are $N$-th roots of unity
such that $\alpha\beta\gamma=1$.

When compactified on $S^1$, the web of 5-branes, or equivalently
M-theory on the non-compact Calabi-Yau,
gives rise at low energy to a 4d field theory.
The 5d vector multiplet gives a 4d vector multiplet, and the real scalar pairs up with the Wilson line along $S^1$ to form a complex scalar.
This is true for both dynamical and background vectors, therefore parameters and moduli of the web end up in parameters and moduli of the 4d theory.
The main proposal of this paper is that the $N$-junction configuration compactified on $S^1$ at low energy gives rise to the $T[A_{N-1}]$ theory constructed in \cite{Gaiotto:2009we}. We devote the rest of this section to perform
various test of this proposal.

Let us first recall the salient properties of  the $T[A_{N-1}]$ theory \cite{Gaiotto:2009we,Gaiotto:2009gz}.
It is a 4d $\cN=2$ isolated SCFT,
obtained by wrapping $N$ M5-branes on a sphere with 3 {\em full punctures},
each of which carries an $\SU(N)$ global symmetry.  Therefore the flavor symmetry
is at least $\SU(N)^3$, with $3(N-1)$ associated mass parameters.
The complex dimension of the Coulomb branch is
\begin{equation}
\label{coulomb dim}
\dim_\bbC \calM_\text{Coulomb}=\frac{(N-1)(N-2)}{2} \;.
\end{equation}
The scaling dimensions of the operators parameterizing the Coulomb branch are $3,4,\ldots,N$
and the multiplicity of the operators of dimension $d$ is $d-2$.
The quaternionic dimension of the Higgs branch can be easily found
from the effective number $n_v$ and $n_h$ of vector- and hypermultiplets calculated in \cite{Gaiotto:2009gz}, and is
\begin{equation}
 \label{higgs dim}
\dim_\bbH \calM_\text{Higgs}= n_h-n_v=\frac{3N^2 - N - 2}{2} \;.
\end{equation}

These are the properties we wish to reproduce from the $N$-junction picture.
First of all notice that the $N$-junction has three copies of $SU(N)$ global symmetry, each realized on the worldvolume of $N$ semi-infinite 5-branes extending in the different directions.
In the next section we will introduce 7-branes on which
semi-infinite 5-branes terminate, without changing the low-energy theory.
Then the global symmetry is realized on the worldvolume of the 7-branes.

\subsection{Coulomb branch}

Let us next count the dimension of the Coulomb branch,
which corresponds to normalizable deformations of the web inside the two-dimensional plane
$(x^5, x^6)$. Deformations of the web are described by real scalars which are in 5d vector multiplets. These are background or dynamical fields
depending on the normalizability of the wave-functions.
Practically it means that a mode is background or dynamical depending on
whether it changes the boundary conditions at infinity.
Each of the single junctions in the web
contributes two real degrees of freedom,
and each of the internal 5-branes establishes one relation between
the positions of the junction points.
We then need to subtract two rigid translations acting on the system as a whole.
Therefore
\be
\label{num deformations}
n_\text{deformations} = 2n_\text{junctions} - n_\text{internal lines} - 2 \;.
\ee

For the $N$-junction configuration,
\begin{equation}
n_\text{junctions}=N^2 \;, \qquad n_\text{internal lines}=\frac32 N(N-1) \;,
\end{equation}
which means
\begin{equation}
n_\text{deformations} = \frac{(N-1)(N+4)}2 \;,
\end{equation}
Each bunch of $N$ semi-infinite 5-branes has $N-1$ non-normalizable deformations which break the $SU(N)$ global symmetry factor and are in correspondence with its Cartan generators.
In the toric diagram, these correspond to points on the edges.
Subtracting them we find
\begin{equation}
\label{Coulomb branch max punct}
\dim_\bbC \calM_\text{Coulomb} =\frac{(N-1)(N-2)}2 \;,
\end{equation}
reproducing \eqref{coulomb dim}.

The dimension of the Coulomb branch can directly be determined as
the number of closed faces in the web diagram -- this will be true even in the
more general configurations introduced in the next sections.

\subsection{Higgs branch}

In order to see the Higgs branch, which corresponds to local deformations as well, we need to terminate the semi-infinite 5-branes on 7-branes at some finite distance.
The same procedure was adopted, for instance, in \cite{DeWolfe:1999hj} to study some 5d conformal theories, or in \cite{Witten:1997sc} (where D4-branes end on D6-branes) to study 4d gauge theories. A semi-infinite D5-brane can end, without breaking any further supersymmetry, on an orthogonal spacetime filling D7-brane, and more generally a $(p,q)$ 5-brane can end on a $[p,q]$ 7-brane as obtained by application of $SL(2,\bbZ)$-duality of type IIB string theory.
The configuration we adopt was shown in Table \ref{tab: web config}
and the 5-brane web is completely suspended between parallel 7-branes.

As analyzed in \cite{Hanany:1996ie} in a T-dual setup, the low energy 5d dynamics on a 5-brane suspended between a 5-brane and a 7-brane does not contain any vector multiplet. The motion of the 7-brane in the direction of the 5-brane is not a parameter of the 5d theory, such that the length of the 5-brane can be taken to infinity recovering the previous setup, or kept finite. On the other hand the motion of the 7-brane orthogonal to the 5-brane is, as before, a non-normalizable deformation.

Once all semi-infinite 5-branes end on 7-branes,
the global symmetries can be seen as explicitly realized on the 7-branes \cite{DeWolfe:1998eu,DeWolfe:1999hj}.
Each of them has a $U(1)$ gauge theory living on its worldvolume. When 7-branes of various type can collapse to a point in the $(x^5,x^6)$-plane,
gauge symmetry enhancement will occur on their worldvolume,
and states of the 5d theory fall naturally under representations of this enhanced symmetry group.
The simplest case is when $k$ 7-branes of the same type collapse to a point.
In a duality frame this just corresponds to $k$ D7-branes at a point,
showing $SU(k)$ flavor symmetry.

The dimension of the Higgs branch is maximal when all parallel 5-branes are coincident, that is when all mass deformations are switched off and the global symmetry is unbroken, and we are at the origin of the Coulomb branch.%
\footnote{Mass deformations reduce the Higgs branch dimension. Moreover along the Coulomb branch various mixed Coulomb-Higgs branches originate, as in more familiar \Nugual{2} theories, see for instance \cite{Argyres:1996eh}. This full structure of the moduli space could be studied as well.}
In this case the central $N$-junction can split into $N$ separate simple junctions, free to move on the $x^{7,8,9}$ plane. The compact component along the $(x^5,x^6)$-plane of the gauge field on the 5-branes pairs up with the three scalars encoding the $x^{7,8,9}$ position to give a hyper-K\"ahler Higgs moduli space. Moreover each bunch of $N$ parallel 5-branes can fractionate on the 7-branes, in the same way as it happens in type IIA \cite{Witten:1997sc}. After removing the decoupled center of mass motion we get the dimension of the Higgs moduli space:
\be
\label{Higgs branch max punct}
\dim_\bbH \calM_\mr{Higgs} = N -1 + 3 \sum_{i=1}^{N-1} i = \frac{3N^2 - N - 2}{2} \;.
\ee
Reassuringly, it agrees with the known value \eqref{higgs dim}.
Issues related to the s-rule will be discussed in section \ref{sec: punctures}.
In section \ref{sec: examples} we will discuss many specific examples.

\subsection{Dualities and Seiberg-Witten curve}

The web of 5-branes in type IIB string theory we consider here can be mapped to different  setups of string or M-theory by various dualities.
For instance, consider a configuration without 7-branes
where all of them has been moved to infinity.
Doing a T-duality along the direction $x^4$ after having compactified it,
we get a system of D6-branes and KK monopoles in type IIA.
This can be further uplifted to M-theory, where everything becomes pure geometry:
a toric conical Calabi-Yau threefold singularity
whose toric diagram is the dual diagram to the 5-brane web.
For the $N$-junction configuration, we have M-theory on $\bbC^3/\bbZ_N \times \bbZ_N$.
Our proposal was that  at low energy this gives a 5d field theory, which after further reduction to 4d flows to the $T[A_{N-1}]$ theory.
The flavor symmetry $\SU(N)^3$ is then realized on the homology of the singularity, by M2-branes wrapping vanishing 2-cycles.
Webs of 5-branes which require mutually non-local 7-branes, of which we will see many examples in Sec.~\ref{sec: examples}, are still expected to be mapped to pure geometry in M-theory, however not to a toric geometry.

Another chain of dualities which we use is the following.
Consider the 5d theory compactified on a circle along $x^4$. Type IIB string theory on a circle is dual to M-theory on a torus. The web of 5-branes is then mapped to a single M5-brane wrapping a holomorphic curve on $\bbC^* \times \bbC^*$. We can then send the IIB circle to zero to obtain the 4d theory.
The M5-brane now wraps a curve on $\bbC\times \bbC^*$.
This chain of dualities is closely related to the one described above:
the fibration of $A_1$ singularity over the curve thus obtained
describes the type IIB mirror of the toric Calabi-Yau singularity
in the type IIA description.
We will use these well-developed techniques to
find the SW curve of the $N$-junction theory
compactified on $S^1$ in Appendix \ref{sec: SW curve},
and thus confirm that it indeed gives the SW curve of the $T[A_{N-1}]$ theory
found in \cite{Gaiotto:2009we} in the suitable limit.

\section{General punctures and the s-rule}
\label{sec: punctures}

\subsection{Classification of punctures}

According to \cite{Gaiotto:2009we,Gaiotto:2009gz} when $N>2$ there are more than one possible kind of punctures in the $A_{N-1}$ $(2,0)$ theory. Wrapping $N$ M5-branes on the sphere with 3 generic punctures  gives rise to an  SCFT, up to some restrictions on the type of punctures \cite{Gaiotto:2009we, Gaiotto:2009gz}.
Since there are no marginal parameters associated to a configuration of three points on a sphere, it is an isolated SCFT, but with a more general global symmetry given by the type of the punctures.
The possible type of punctures in the $A_{N-1}$ $(2,0)$ theory are classified by Young tableaux with $N$ boxes. We will see below that such classification naturally arises in our construction. For that purpose, it is sufficient to consider
a bunch of $N$ semi-infinite 5-branes extending in the same direction,
because each of the three bunches corresponds to each of the punctures.

Generically, instead of ending each 5-brane on a different 7-brane, we can group some 5-branes and end them together on the same 7-brane,%
\footnote{We thank D.~Gaiotto for suggesting this possibility to us.}
which requires the 5-branes to overlap, and  reduces the number of mass deformations according to the fact that the flavor symmetry carried by the bunch gets reduced.
Given a bunch of $N$ 5-branes, we can group them according to a partition $\{k_i\}$ of $N$  with $\sum_i k_i = N$, and end $k_i$ of them on the $i$-th 7-brane.
Thus the possible kind of punctures are naturally classified by partitions.
Partitions can then be represented by Young tableaux, reproducing the classification in \cite{Gaiotto:2009we}.
A similar construction involving D3- and D5-branes was employed in \cite{Gaiotto:2008sa} to understand the possible boundary conditions of $\cN=4$ super Yang-Mills theory.

A set of $n$ bunches made of the same number $k$ of 5-branes carries a $U(n)$ flavor group. However a diagonal $U(1)$ for the whole set of $N$ 5-branes is not realized on the low energy theory \cite{DeWolfe:1999hj}. Then the flavor symmetry of the puncture is $S\big( \prod_k U(n_k) \big)$, where $n_k$ is the number of bunches of $k$ 5-branes. This agrees with the flavor symmetry associated to a puncture, found in \cite{Gaiotto:2009we}. The full puncture corresponds to the partition%
\footnote{Here and in the following, with the notation $\{A^b\}$ we mean the partition $\{\underbrace{A, \dots , A}_{b \text{ times}} \}$.}  $\{1^N\}$.

It is easy to count the dimension of the Higgs branch for an arbitrary choice of 3 punctures. The internal web always contributes $N-1$ (decoupling the center of mass). Each puncture contributes according to its defining partition  $\{k_i\}_{i=1\dots J}$. Let us conventionally order $k_1 \geq \dots \geq k_J$, then the counting of legs gives for the Higgs branch at the puncture $\calM_H^p$:
\be
\dim_\bbH \calM_H^p = \sum_{i=1}^J (i-1) \, k_i = -N + \sum_{i=1}^J i\, k_i \;.
\ee
It is easy to check that for the partition $\{1^N\}$ we get one of the three terms in (\ref{Higgs branch max punct}). Many examples and comparisons with known results are in Section \ref{sec: examples}.
In order to count the dimension of the Coulomb branch, we need a precise understanding of the s-rule, to which we devote the next subsection.

It is worth stressing how this construction makes it clear that a theory with punctures of a lower type is effectively embedded into the Higgs branch of a theory with only punctures of the maximal type, \ie{} obtained using the maximal number of 7-branes.
We saw that when two or more 7-branes of a puncture are aligned in such a way that the parallel 5-branes ending on them overlap, we can break the 5-branes on the 7-branes and move the cut pieces around, to realize Higgs branches. When the extra pieces are taken very far away, \ie{} when one gives large VEV's and goes to the Higgs branch, they effectively decouple from the rest of the web, and some 7-branes can be left effectively disconnected. One is left with a puncture composed of a smaller number of 7-branes, and multiple 5-branes ending on the same 7-branes, that is a more generic puncture. This shows that the effective theory along the Higgs branch under consideration is the SCFT related to the puncture of ``lower type'', plus some decoupled modes describing the motion of the extra 5-brane pieces.

\subsection{Generalized s-rule}
\label{sec: s-rule dot diagram}

In general the Coulomb branch gets reduced by lowering the degree of the punctures. This is due to the \emph{s-rule}, which was originally introduced in \cite{Hanany:1996ie} in the construction of 3d gauge theories in order to correctly account for the dimension of mixed Coulomb-Higgs branches when D3-branes end on D5's and NS5's, and later studied in \eg{} \cite{Ooguri:1997ih, Bachas:1997kn, Bachas:1997sc, Hori:1997ab}.

The s-rule states that there are no supersymmetric states if more than one D3-brane is suspended between a given pair of D5 and NS5.
The same rule is necessary to correctly describe the dynamics of D4-branes between D6's and NS5's, see for instance \cite{Elitzur:1997fh}.
In simple cases it would be enough for us to use a T-dual version of it, that is, we cannot have more than one D5-brane between a D7 and an NS5.
Any $SL(2,\bbZ)$-dual version of this statement is also an s-rule.
However we need a version of the s-rule which applies to
general intersections of $(p,q)$ 5-branes suspended between different numbers of 7-branes.
This question was answered in \cite{Mikhailov:1998bx,DeWolfe:1998bi,Bergman:1998ej} in the context of string junctions in the presence of 7-branes,
which we can directly borrow because the supersymmetry conditions
on space-filling 5-brane and string junctions are essentially the same.
We will also need to understand how the s-rule propagates inside complicated 5-brane webs.
Both issues, carefully described below, can be understood using
the brane creation effect \cite{Hanany:1996ie}
when a 7-brane crosses a 5-brane, and are explained in section \ref{sec: generalized-s-rule}.

For the sake of clarity, we prefer to state the rule,
leaving any derivation to the next subsection.
The s-rule is better visualized on the diagram dual to the web: we call it a \emph{dot diagram} instead of a toric
diagram, since in the general case it does not represent a toric geometry.
Given a web of 5-branes which do not require 7-branes, the dot diagram is constructed on a square lattice by associating a dot to each face (even non-compact) in the web, and a line connecting two dots whenever the two faces are adjacent. The lines in the dot diagram must be orthogonal to the 5-branes in the web. It is always possible to go  back and forth from the web diagram to the dot diagram, by exchanging 5-branes with orthogonal lines and vice versa.
Notice that the web diagram encodes the parameters and moduli of the 5d field theory, whilst all this information is lost in the dot diagram.
The boundary conditions in the web determine the external lines in the dot diagram, which form a convex polygon, whereas the details of the web determine a tessellation of such polygon.
In this particular case, the dot diagram is really a toric diagram. Moreover it is completely triangulated by minimal triangles of area $1/2$; this is because in the web all junctions are trivalent.

In the presence of 7-branes, we proceed as follows.
\begin{itemize}
\item In the dot diagram, we can represent the fact that $n$ parallel 5-branes end on the same 7-brane by separating $n-1$ consecutive segments by a white dot, as opposed to a black one. These $n-1$ segments act as one edge of the minimal polygons, defined momentarily.
We say that these segments {\em bear an s-rule}, in the sense that supersymmetric configurations are now constrained. The boundary conditions in the web determine a convex polygon, made of the external lines in the dot diagram. Consecutive segments can be separated by white or black dots, depending on whether the corresponding parallel 5-branes end on the same 7-brane or not, respectively;
5-branes which do not end on the same 7-brane turn into segments separated by black dots in the dot diagram.
\item Then we proceed to tessellate the dot diagram with minimal polygons.
Consecutive segments separated by a white dot act as a single edge of a minimal polygon. A minimal polygon can be either a triangle or a trapezium, and it must satisfy an extra constraint:
    \begin{itemize}
    \item If it is a triangle, the three edges must be composed by the same number, say $n$, of collinear lines.
    \item If it is a trapezium, there must be two integers $n_1 < n_2$ such that the four edges are made of $n_2$, $n_1$, $n_2-n_1$, $n_1$ segments; furthermore the edges with $n_2$ and $n_2- n_1$ segments must be parallel.
    \end{itemize}
\item In general the tessellation of the dot diagram leads to internal consecutive segments which are separated by white dots, and they again act as a single edge of minimal polygons. This is a \emph{propagation of the s-rule} inside the dot diagram, and has to be respected.
\end{itemize}

\begin{figure}[tn]
\begin{center}
\includegraphics[width=\textwidth]{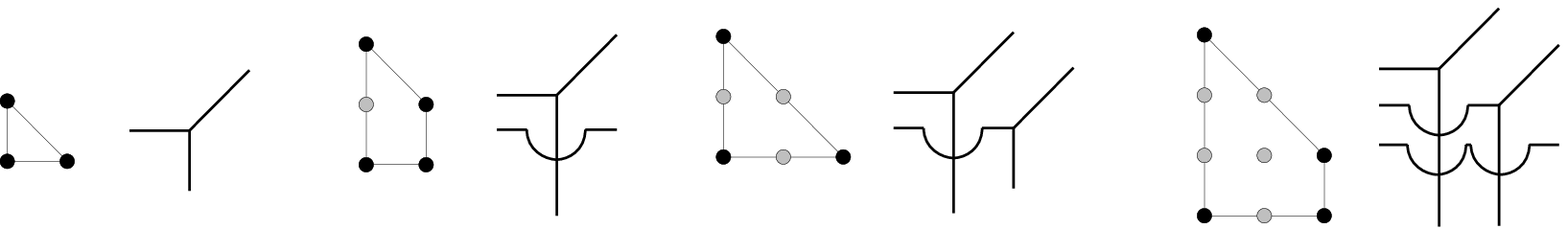}

\vspace{3ex}

\includegraphics[width=.72\textwidth]{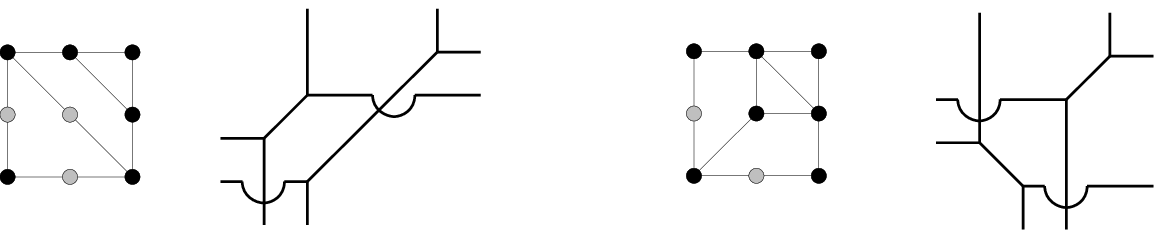}
\end{center}
\caption{\small Upper line: examples of minimal polygons with the dual 5-brane web. White dots separate consecutive segments which act as a single edge. They represent s-rules, and come from either multiple 5-branes ending on the same 7-brane, or propagation of the s-rule inside. In the web, some 5-branes {\em jump over} another 5-brane,
meaning that they cross it without ending.
Lower line: two examples of different allowed tessellations of a $2\times 2$ square, given the same constraints on the external edges. The webs of five-branes are related by a ``flop transition''.
\label{fig: s-rule}}
\end{figure}

If no consistent tessellation exists, it means that the web has no SUSY vacuum and some 7-branes have to be added. Notice that for boundary conditions such that each 5-brane ends on its own 7-brane, the prescription gives back a complete triangulation of the dot diagram in terms of area $1/2$ triangles and only black dots.

Once a consistent tessellation of the dot diagram has been found, we can go back to the web diagram mapping lines to orthogonal 5-branes. Area $1/2$ triangles are mapped to the usual junction of three 5-branes. The other minimal polygons are mapped to intersections of 5-branes in which, because of the s-rule, a 5-brane cannot terminate on another one and has to cross it. We say that the 5-brane {\em jumps over}  the other one, even though there is no real displacement. In figure \ref{fig: s-rule} there are some examples of minimal polygons with the dual web.
For instance, a triangle of edge $n$ is mapped to the intersection of three bunches of $n$ parallel 5-branes in which, because of the s-rule, only $n$ trivalent junctions can occur. A trapezium of edges $n_2$ and $n_1$ is mapped to the intersection of four bunches of $n_2$, $n_1$, $n_2-n_1$, $n_1$ 5-branes in which only $n_1$ junctions can occur, and $n_2-n_1$ 5-branes simply go straight crossing everything.

Let us stress that when multiple parallel 5-branes end on the same 7-brane, they have to be coincident. As we saw this is represented by white dots in the dot diagram. When the s-rule propagates and there are white dots inside, the corresponding 5-branes have to be coincident as well, simply because of the geometric constraint. Notice that tessellations respecting the s-rule are not unique, just as they are not in the unconstrained case: different tessellations are related by changing parameters or moving along the Coulomb branch. One can finally check that this prescription agrees with the usual s-rule when applicable.

When a 5-brane cannot end on another one and therefore
just crosses it, it can happen that a face in the 5-brane web gets frozen and does no
longer have a modulus related to its size.
This visually shows the effect of the Higgs mechanism:
it gives mass to scalars in vector multiplets which parameterize the size of internal faces,
and so effectively freezes some moduli.
Indeed it can be checked that open faces, as the result of 5-branes jumping over other ones,
geometrically do not have moduli related to their size: they are completely
fixed by the structure of the web. Only (and all) closed faces have one modulus
controlling their size.
The dimension of the Coulomb branch is then easily counted from the web of 5-branes: it is the number of closed internal faces.
We will see many examples in the next section.
Example of consistent tessellations can be found in figures \ref{fig: E7}, \ref{fig: E8}, \ref{fig: E6 rank N}, \ref{fig: E7 rank N}, \ref{fig: E8 rank N}.

\subsection{Derivation of the generalized s-rule}
\label{sec: generalized-s-rule}

Let us now derive the generalized s-rule stated in the previous subsection.
Mostly the same rule was formulated by  \cite{Mikhailov:1998bx,DeWolfe:1998bi,Bergman:1998ej} in the case of the web of $(p,q)$-strings, and we shall soon see that
the same rule applies to the web of $(p,q)$ 5-branes.
We will also emphasize the {\em propagation of the s-rule}
which was not clearly mentioned in the previous literature.

We follow the derivation given by \cite{Bergman:1998ej},
which used the brane creation/annihilation mechanism of \cite{Hanany:1996ie}.
Let us start by reviewing  how the brane creation mechanism works when a 7-brane crosses a 5-brane.
A $[p,q]$ 7-brane creates an $SL(2,\bbZ)$ monodromy $X_{p,q}$ for the axiodilaton $\tau$ given by
\begin{equation}
\label{monodromy}
X_{p,q}= \begin{pmatrix} a & b \\ c & d \end{pmatrix}
= \begin{pmatrix} 1+pq & -p^2 \\ q^2 & 1-pq \end{pmatrix}
\end{equation}
which, following the conventions of \cite{DeWolfe:1998eu}, is measured counterclockwise.
We represent the monodromy as a cut originating from the 7-brane.
Then $\tau$ is transformed as
\begin{equation}
\tau \to  \frac{a\tau + b}{c\tau + d}
\end{equation} when we cross the cut.
Accordingly,  when a 5-brane crosses a branch cut,
it is generically transformed by the monodromy and it changes its slope
in the diagrams we show. This is a schematic way to
depict the correct situation in the true curved geometry,
in which the 5-brane just follows a geodesic. In all our web constructions we choose the cut in such a way that they do not intersect the web.

\begin{figure}[tn]
\centerline{
\includegraphics[height=10ex]{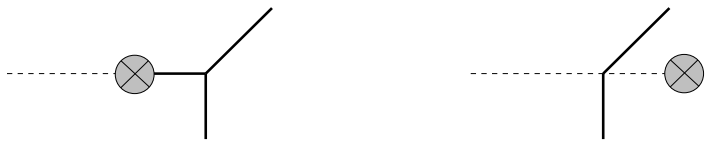}
}
\caption{\small brane creation mechanism for 5-branes. On the left an NS5 becomes a $(1,1)$ 5-brane because it meets a D5, which comes from a D7-brane shown by a $\otimes$ sign. The dotted line shows the cut associated to the monodromy. On the right the D5 has disappeared because of the brane creation/annihilation mechanism, but the boundary conditions are the same as before because of the cut, along which $\tau \to \tau - 1$.
\label{HananyWitten1}}
\end{figure}

Consider first the usual junction between a D5, an NS5 and a $(1,1)$ 5-brane, and suppose the D5 ends on a D7. Let us take the cut to run away without crossing the 5-branes. We can then move the D7 to the other side of the NS5: when they cross the D5 disappears by the brane creation/annihilation mechanism, however the NS5 now crosses the branch cut. We are left with an NS5 which becomes a $(1,1)$ 5-brane when crossing the cut. The process is shown in figure \ref{HananyWitten1}.

\begin{figure}[tn]
\centerline{
\includegraphics[width=\textwidth]{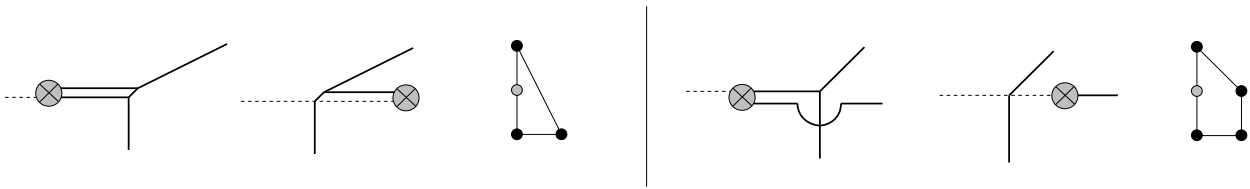}
}
\caption{\small Left: non-SUSY configuration, which violates the s-rule. SUSY breaking is apparent in the second figure, where an anti-D5 is present. The polygon does not respect the s-rule either. Right: SUSY configuration. Only one D5 ends on the NS5. The polygon is acceptable.
The 5-brane which cannot end on the other one and therefore just crosses it was shown as if it jumps over the other one.
\label{HananyWitten2}}
\end{figure}

It is easy to understand the s-rule in its standard formulation: in a SUSY configuration, no more than one D5-brane can stretch between a D7 and an NS5. See figure \ref{HananyWitten2}.
Suppose we cook up such forbidden configuration.
When moving the D7 to the other side of the NS5, all but one D5's remain as 5-branes stretched between the D7 and the NS5.
Charge conservation at the junction requires them to be anti-D5's, showing that SUSY must be broken; moreover tensions do not balance anymore.
On the other hand, if only one D5 ends on the NS5 while all other ones cross without terminating, the configuration is still supersymmetric after pulling the D7 to the other side.

In figure \ref{HananyWitten2} we also showed the dual dot diagrams, whose precise construction has been given above. It is easy to check that dot diagrams corresponding to non-supersymmetric configurations, do not respect the s-rule prescription we gave.

\begin{figure}[tn]
\centerline{
\includegraphics[width=\textwidth]{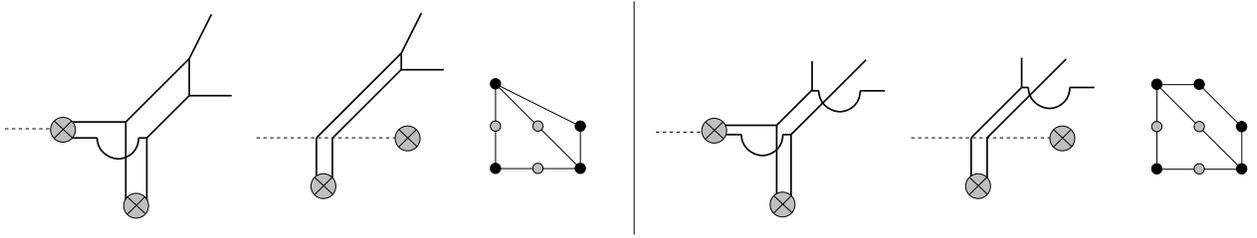}
}
\caption{\small propagation of the s-rule. Left: non-SUSY configuration. Even though the s-rule is respected where the D5's meet the NS5's, it is violated where both $(1,1)$ 5-branes meet the same 5-brane. The propagation of the s-rule is manifest in the second figure. The polygon violates the s-rule as well. Right: a SUSY configuration, with the corresponding polygon.
\label{HananyWitten3}}
\end{figure}

The generalization to more involved configurations is straightforward.
In particular let us show that the s-rule {\em propagates},
see Figure \ref{HananyWitten3}.
Consider a configuration where two D5-branes, ending on the same D7, meet two NS5-branes, ending on the same $[0,1]$ 7-brane.
Each D5 can end on a different NS5, resulting in two $(1,1)$ 5-branes. This is as in the previous example.
The novelty is that the two $(1,1)$ 5-branes still carry a constraint: they behave
as if they came from the same $[1,1]$ 7-brane -- in particular they cannot end on the same 5-brane.

In order to see why, let us  pull the D7 to the other side of the NS5's. We are left with two NS5-branes, ending on the same $[0,1]$ 7-brane, which become two $(1,1)$ 5-branes when crossing the cut. Now the usual
s-rule applies, in a different S-dual frame.
Namely, there cannot be two NS5-branes stretching between a D5 and a $[0,1]$ 7-brane.
In figure \ref{HananyWitten3} we also showed the dual dot diagrams with tessellation, one allowed and one not. Generalizing these examples, one gets the set of rules
we stated in the previous subsection.

\section{Examples}
\label{sec: examples}

In this section we consider many examples of increasing complexity, where all rules previously stated will become clear. Comparisons with known field theories will be made when possible.

\subsection{$N=2$}
\label{sec: N=2}

The 4d low energy theory on the $N=2$ multi-junction has $SU(2)^3$ global symmetry, three mass deformations corresponding to the Cartan generators of the flavor group, $\dim_\bbC \calM_C =0$ and $\dim_\bbH \calM_H = 4$.
This is the theory $T[A_{1}]$;
in fact, it is given by 8 free chiral superfields $Q_{ijk}$, where each index is in the $\rep{2}$ of one $SU(2)$.

\begin{figure}[tn]
\centerline{
\begin{tabular}{cccc}
a) generic & b) $m_1=\pm m_2\pm m_3$ & c) $m_1=0$, $m_{2,3}$ generic & d)  $m_1=0$, $m_2=\pm m_3$\\[1em]
\includegraphics[height=16ex]{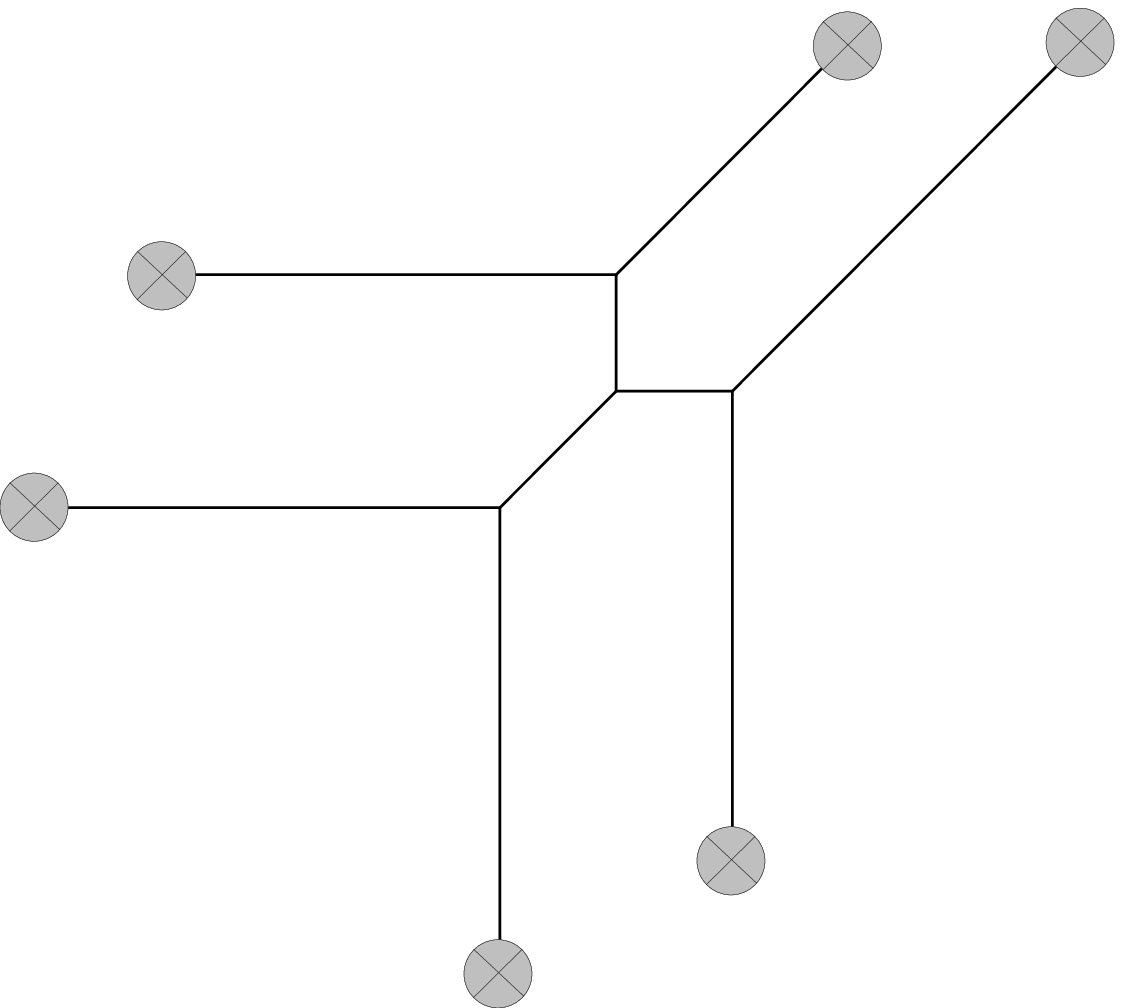}
&
\includegraphics[height=15ex]{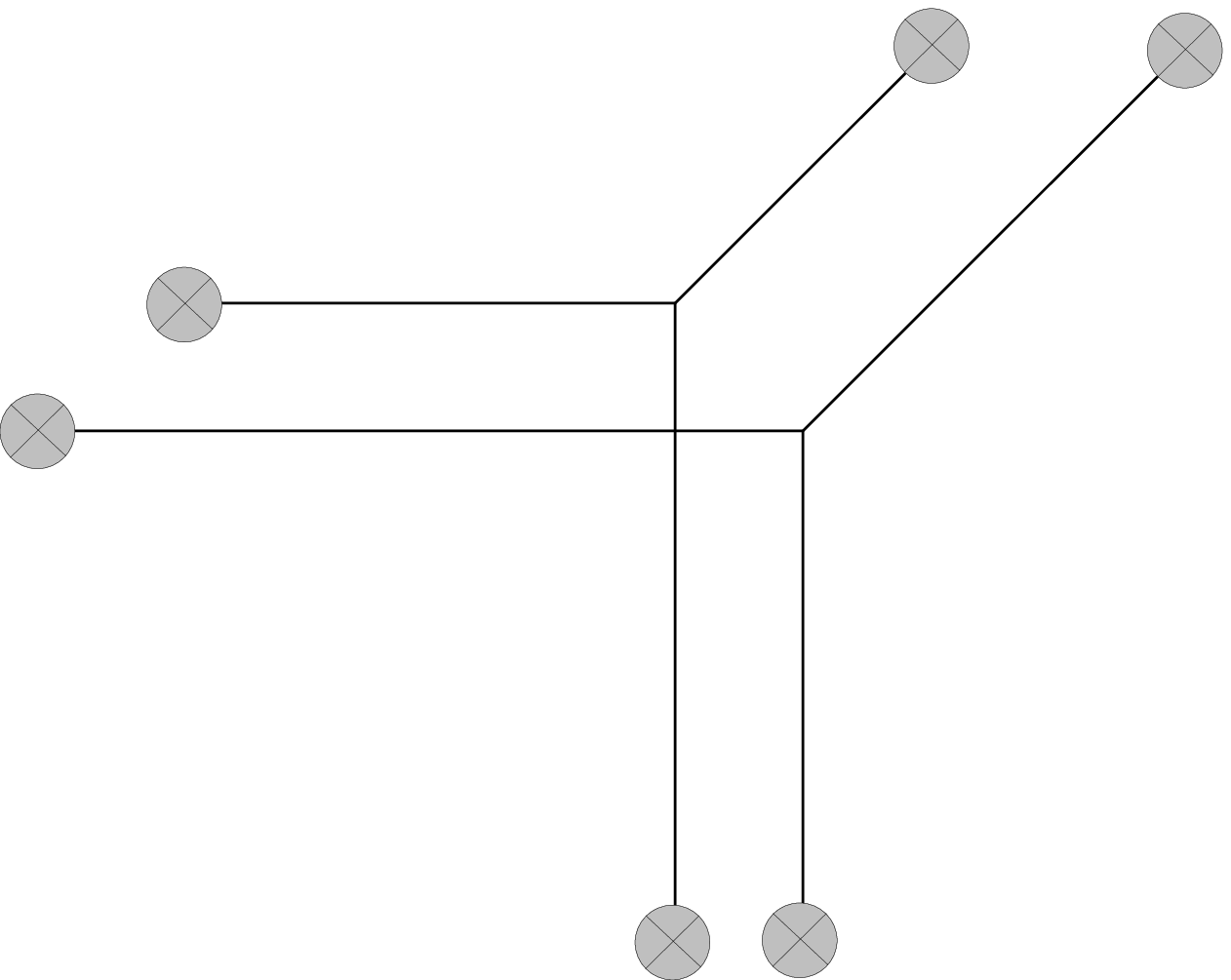}
&
\includegraphics[height=15ex]{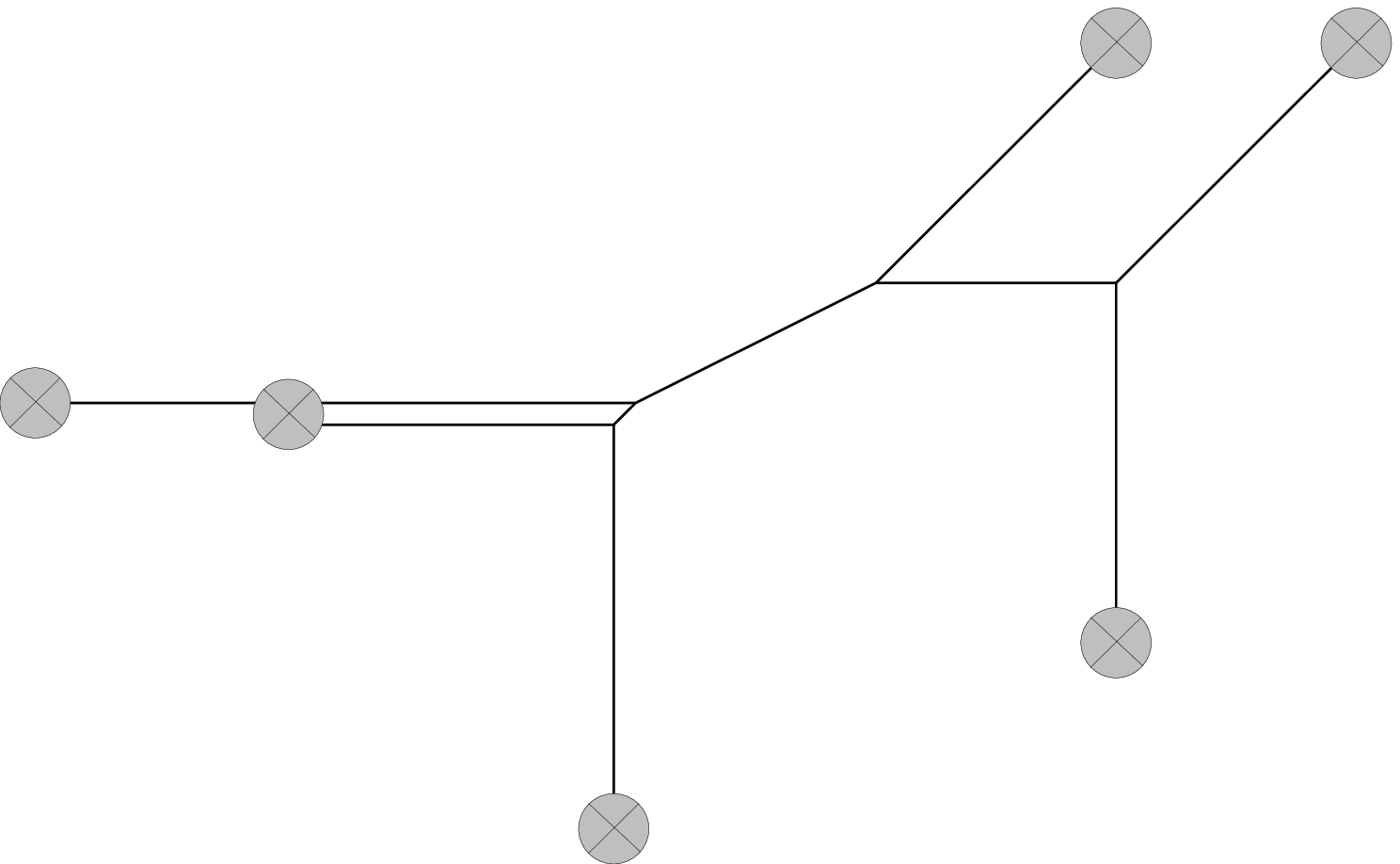}
&
\includegraphics[height=13ex]{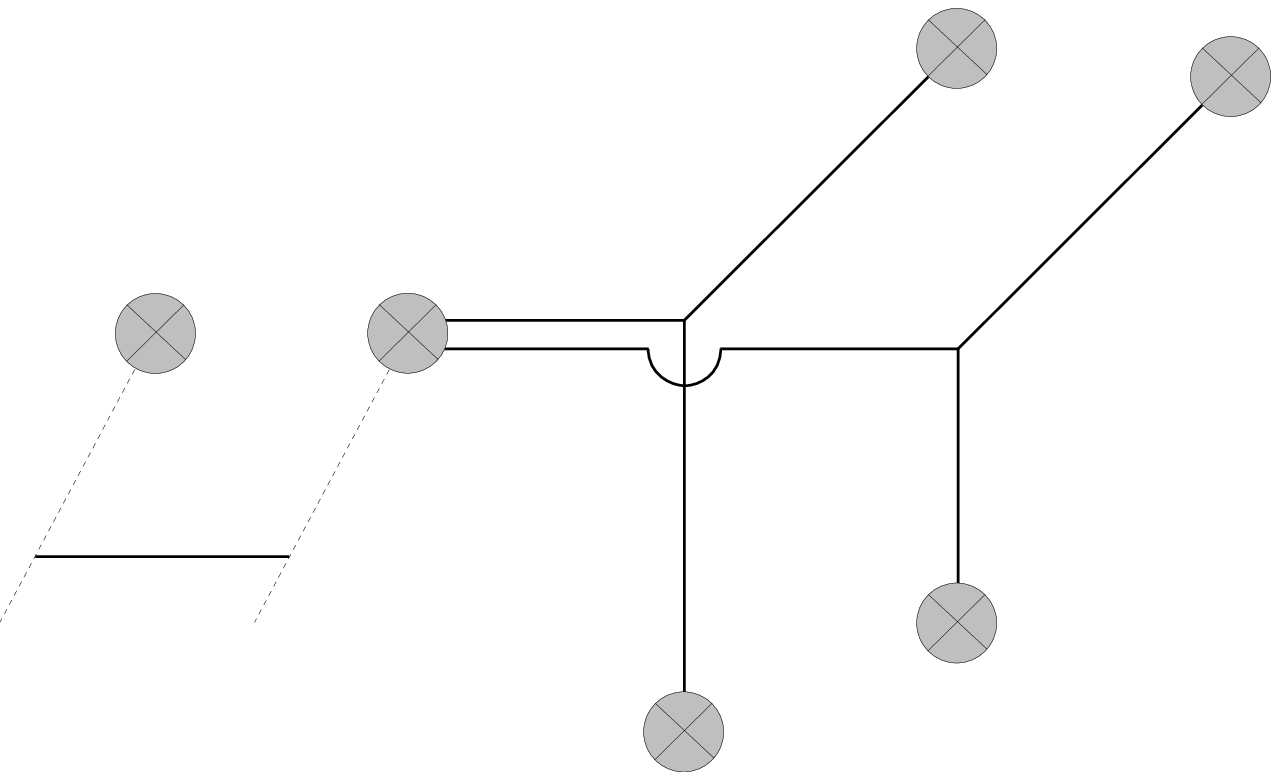}
\end{tabular}
}
\caption{\small Higgs branches of the $N=2$ multi-junction for various values of the mass parameters. From left to right. a) Generic masses and no Higgs branch. b) $m_1 = \pm m_2 \pm m_3$ and one Higgs direction corresponding to splitting the multi-junction in two simple junctions. c) $m_1=0$ and no Higgs branch due to the s-rule. The piece of D5 on the left cannot be removed. d) $m_1=0$ and $m_2 = \pm m_3$, now one D5 can jump the NS5, the s-rule is satisfied and the piece of D5 can be removed, as well as the junction can be split.
\label{fig: N=2}}
\end{figure}

Even though trivial, this system allows us to perform a nice check of the s-rule.
The superpotential for general mass deformations  is
\be
W = Q_{ijk} Q_{lmn} (m_1 \delta^{il} \epsilon^{jm} \epsilon^{kn} + m_2 \epsilon^{il} \delta^{jm} \epsilon^{kn} + m_3 \epsilon^{il} \epsilon^{jm} \delta^{kn} ) \;
\ee
where $m_{i}$ is the mass parameter associated to the $i$-th $SU(2)$
flavor symmetry.
Diagonalizing the mass matrix, the hypermultiplet masses are then  $\pm m_1\pm m_2\pm m_3$.
The dimension of the Higgs branch is the number of the massless hypermultiplets.
Let us reproduce this from the s-rule:
\begin{itemize}
\item If the three masses are generically non-zero and the flavor group is broken to $U(1)^3$ there is no Higgs branch. If the masses satisfy $m_1 = \pm m_2 \pm m_3$ there is a single Higgs branch direction. In the web of 5-branes this corresponds to aligning the 7-branes such that the multi-junction can split in two simple junctions, see Figure \ref{fig: N=2}.
\item If one mass is zero and the other two generic, there is no Higgs branch. In the web of 5-branes this is guaranteed by the s-rule, see figure \ref{fig: N=2}. If the two masses further satisfy $m_2 = \pm m_3$ there is a two-dimensional Higgs branch, corresponding to removing one piece of 5-brane and splitting the remaining web in two simple junctions.
\item If all masses are zero, all four Higgs branch directions open up.
\end{itemize}

\subsection{$N=3$ and the $E_6$ theory}

\begin{figure}
\[
\inc{.3}{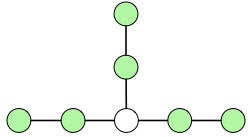}
\]
\caption{\small Extended Dynkin diagram of $E_6$ showing the $\SU(3)^3$ subgroup.
\label{tac:E6dynkin}}
\end{figure}

The $N=3$ multi-junction was already shown in Figure \ref{fig: junctions}.
It has $SU(3)^3$ global symmetry, 6 mass deformations and then no marginal couplings, $\dim_\bbC \calM_C = 1$ and $\dim_\bbH \calM_H = 11$.
It has already been noticed in \cite{Gaiotto:2009we} that for $N=3$ the visible $SU(3)^3$ global symmetry is actually enhanced to $E_6$.
We see here that even the Higgs branch dimension works out correctly:
indeed, the Higgs branch is the centered one-instanton moduli space of $E_6$,
whose quaternionic dimension is $11$.
Some properties of the exceptional SCFT's are collected in appendix \ref{sec: exp theories}.
In fact, $\bbC^3/\bbZ_N \times \bbZ_N$ is described by the equation $xyz=t^N$ in $\bbC^4$. For $N=3$ it is an homogeneous cubic equation, and thus a complex cone over its projection. Its projection gives a cubic in $\bbP^3$, which is the del Pezzo surface dP$_6$ at a particular point of its complex structure moduli space where it is toric.
Its homology realizes the lattice of $E_6$, and thus M-theory on the CY$_3$ cone over dP$_6$ has $E_6$ global symmetry in 5d.

Another way to understand the symmetry enhancement is to study the monodromy
of the system of 7-branes. Let us recall that the $SL(2,\bZ)$ monodromy
around a $[p,q]$ 7-brane is given by \eqref{monodromy}.
Let us denote the monodromy matrices of the 7-branes we use as
\begin{equation}
P=X_{1,0} \;,\qquad
Q=X_{0,1} \;,\qquad
R=X_{1,1} \;.
\end{equation}
In the literature, the basis is usually given instead by
\begin{equation}
A=X_{1,0} \;,\qquad
B=X_{1,-1} \;,\qquad
C=X_{1,1}
\end{equation}
so that a single O7-plane splits into $CB$ non-perturbatively.
The combined monodromy of our system is then $R^3 Q^3 P^3$,
which is conjugate to $(CB)^2 A^5$, known as the affine $E_6$ configuration.%
\footnote{Following the conventions of \cite{DeWolfe:1998eu}, the 7-branes are an ordered system according to the order we meet their cuts circling counterclockwise. Obviously, the corresponding monodromies have to be multiplied in the opposite order. Here and in the following we always report the monodromy matrices.}
It is known that eight out of nine 7-branes can be collapsed together,
making the F-theory 7-brane of type $E_6$. Our configuration shows instead that
we can collapse three bunches of three 7-branes,
displaying the $\SU(3)^3$ subgroup of $E_6$, see Fig.~\ref{tac:E6dynkin}.
More arguments supporting this identification will be presented in Sec.~\ref{sec: davide}.

We can realize the other puncture, represented by the partition $\{1,2\}$ and giving rise to $U(1)$ flavor symmetry, from the $\{1,1,1\}$ puncture by moving away 2 pieces of 5-brane along the Higgs branch. Performing this operation on one of the three punctures, we get a theory with $SU(3)^2 \times U(1)$ symmetry, $\dim_\bbC \calM_C = 0$ and $\dim_\bbH \calM_H = 9$. The  Coulomb branch is lifted by the s-rule. This theory corresponds to 9 free hypermultiplets $Q_{ij}$, each index being in the $\rep{3}$ of one $SU(3)$.

\subsection{$N=4$ and the $E_7$ theory}

The $N=4$ multi-junction has $SU(4)^3$ global symmetry, 9 mass deformations, $\dim_\bbC \calM_C = 3$ and $\dim_\bbH \calM_H = 21$.
This is the theory $T[A_3]$.

\begin{figure}[tn]
\centerline{
\hspace{\stretch{1}}
\includegraphics[height=26ex]{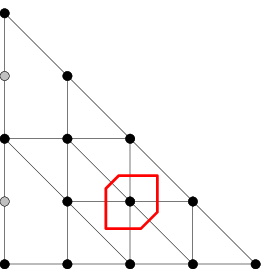}
\hspace{\stretch{2}}
\includegraphics[height=26ex]{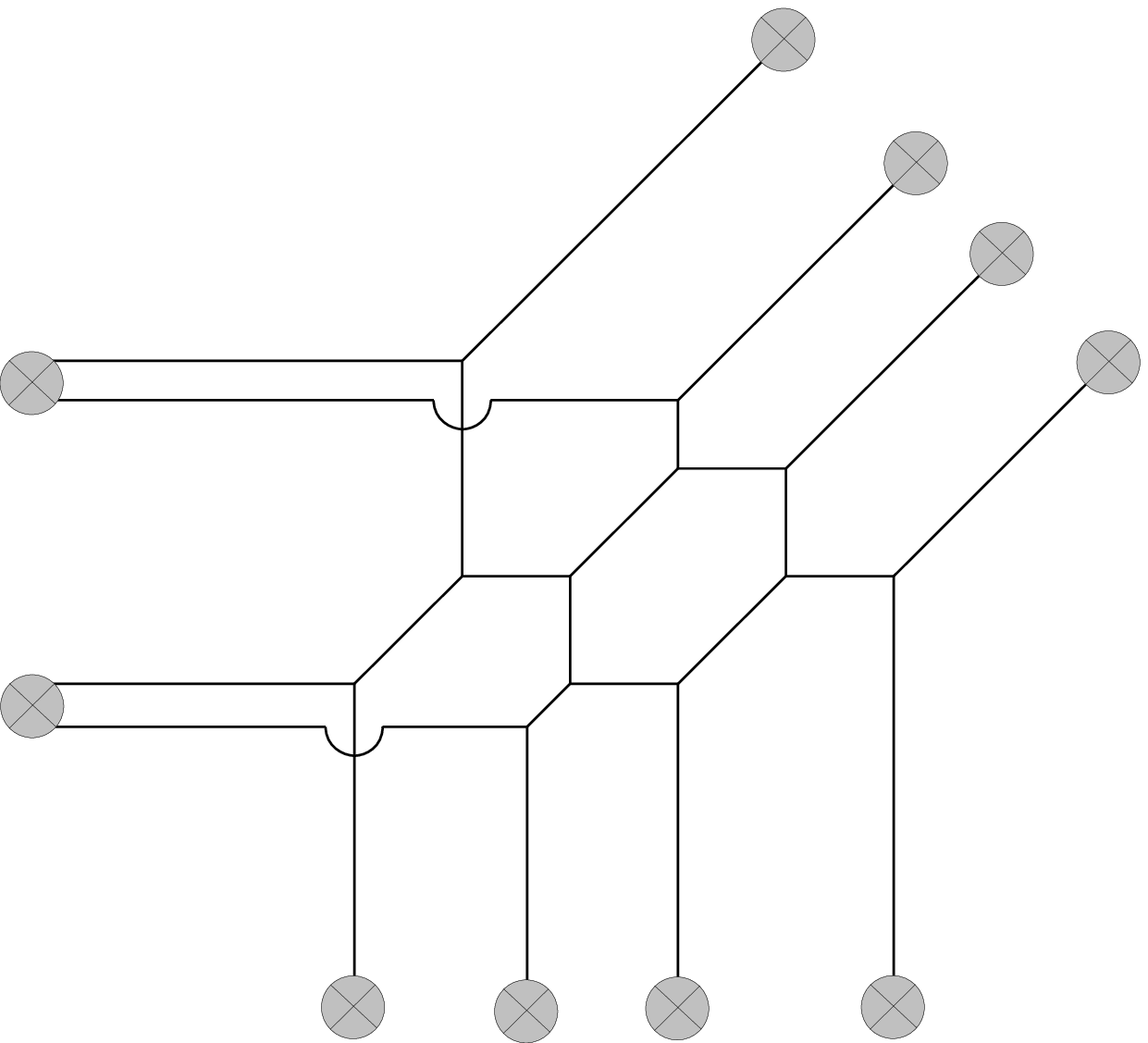}
\hspace{\stretch{1}}
}
\caption{\small Web of 5-branes for the $E_7$ theory. On the left dot diagram of $\bbC^3/\bbZ_4 \times \bbZ_4$ where the tessellation realizes, on the external edges, the partitions $\{1,1,1,1\}$ and $\{2,2\}$. White dots on the edges separate collinear lines that have to be thought of as a single line.
In red the only closed dual polygon, which is a closed face in the web. On the right, dual web with jumps corresponding to the tessellation. The visible symmetry is $SU(4)^2 \times SU(2)$. The only closed face is well visible.
\label{fig: E7}}
\end{figure}

\begin{figure}
\[
\inc{.3}{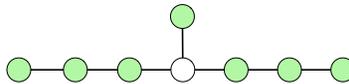}
\]
\caption{\small Extended Dynkin diagram of $E_7$ showing the $\SU(2)\times \SU(4)^2$ subgroup.
\label{tac:E7dynkin}}
\end{figure}

It is particularly interesting to realize the $SU(4)^2 \times SU(2)$ theory, using punctures of partition $\{1^4\}$  and $\{2^2\}$.
It has $\dim_\bbC \calM_C = 1$ and $\dim_\bbH \calM_H = 17$, see Figure \ref{fig: E7} for its 5-brane web and application of the s-rule.
The rank-7 global symmetry is believed to enhance to $E_7$.
Indeed, the Higgs branch has the correct dimension as the one-instanton moduli space of $E_7$.
Let us study the monodromy as we did for $E_6$. The total monodromy is
$R^2 Q^4 P^4$, which is conjugate to $(CB)^2 A^6$ that is known as the affine $E_7$ configuration
of the 7-brane. Nine out of ten 7-branes can be collapsed to one point, making
an F-theory 7-brane of type $E_7$. In our description we instead grouped
four, four and two 7-branes together, showing an $\SU(4)^2\times \SU(2)$ subgroup,
see Fig.~\ref{tac:E7dynkin}.
Notice that the dual geometry we have to compactify M-theory on is non-toric, because we cannot remove the 7-branes without changing the boundary conditions. In fact the 7-th del Pezzo does not have toric points in its complex structure moduli space.
Again, more support for this identification is given in Sec.~\ref{sec: davide}.

\subsection{$N=6$ and the $E_8$ theory}

The $N=6$ multi-junction has $SU(6)^3$ global symmetry, $\dim_\bbC \calM_C = 10$ and $\dim_\bbH \calM_H = 50$. This is the theory $T[A_5]$.

\begin{figure}[tn]
\centerline{
\hspace{\stretch{1}}
\includegraphics[height=30ex]{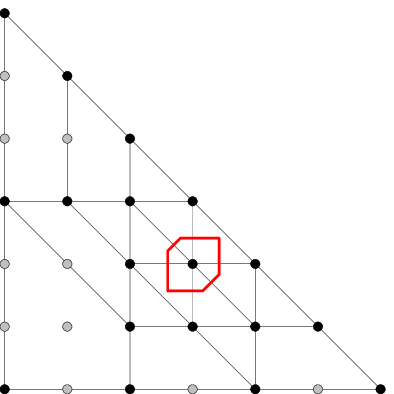}
\hspace{\stretch{2}}
\includegraphics[height=33ex]{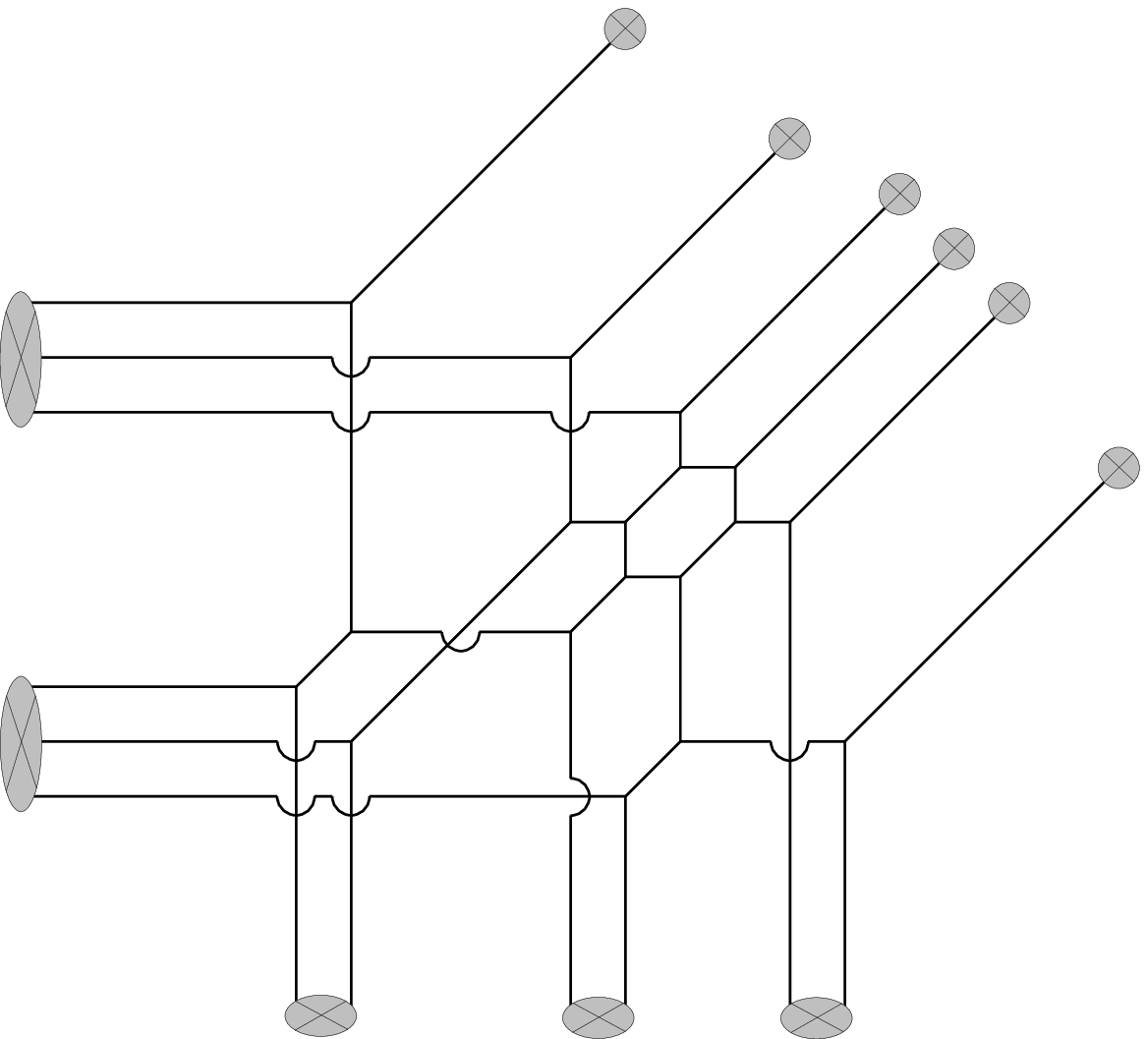}
\hspace{\stretch{1}}
}
\caption{\small Web of 5-branes representing  the $E_8$ theory. On the left dot diagram of $\bbC^3/\bbZ_6 \times \bbZ_6$ where the tessellation realizes the partitions $\{1^6\}$, $\{2^3\}$ and $\{3^2\}$. In red the only closed dual polygon. On the right, dual web with jumps corresponding to the tessellation. The visible symmetry is $SU(6) \times SU(3) \times SU(2)$. The only closed face is well visible.
\label{fig: E8}}
\end{figure}

\begin{figure}
\[
\inc{.3}{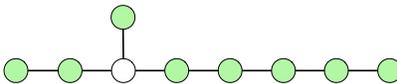}
\]
\caption{\small Extended Dynkin diagram of $E_8$, showing the $\SU(2)\times \SU(3)\times \SU(6)$ subgroup.
\label{tac:E8dynkin}}
\end{figure}

Note that $E_8 \supset SU(6) \times SU(3) \times SU(2)$.
Inspection of partitions shows that $\{2^3\}$ realizes $SU(3)$ and $\{3^2\}$ realizes $SU(2)$. Thus one suspects that the theory $T[A_5]$
contains the $E_8$ theory in its Higgs branch, specified by the three partitions $\{2^3\}$,
$\{3^2\}$ and $\{1^6\}$.
The Coulomb and Higgs branch dimensions can be found from the
5-brane construction,
and are $\dim_\bbC \calM_C = 1$ and $\dim_\bbH \calM_H = 29$,
see Figure \ref{fig: E8}.
These numbers match those of the $E_8$ theory.

Let us study the monodromy produced by the 7-branes. This is now $R^2 Q^3 P^6$,
which is conjugate to $(CB)^2 A^7$ known as the affine $E_8$ configuration.
It is known that ten out of eleven 7-branes can be collapsed to a point,
giving us an F-theory 7-brane of type $E_8$. Instead in our description
an $\SU(2)\times \SU(3)\times \SU(6)$ flavor symmetry is manifest,
see Fig.~\ref{tac:E8dynkin}.

\subsection{Higher-rank $E_n$ theories}

\begin{figure}[tn]
\centerline{
\hspace{\stretch{1}}
\includegraphics[height=30ex]{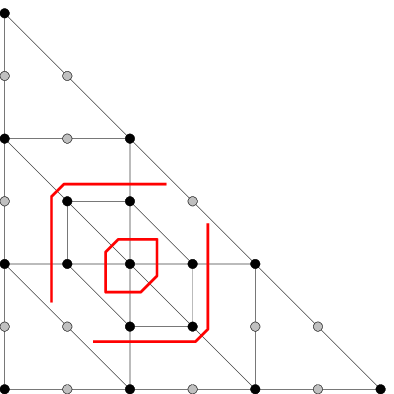}
\hspace{\stretch{2}}
\includegraphics[height=33ex]{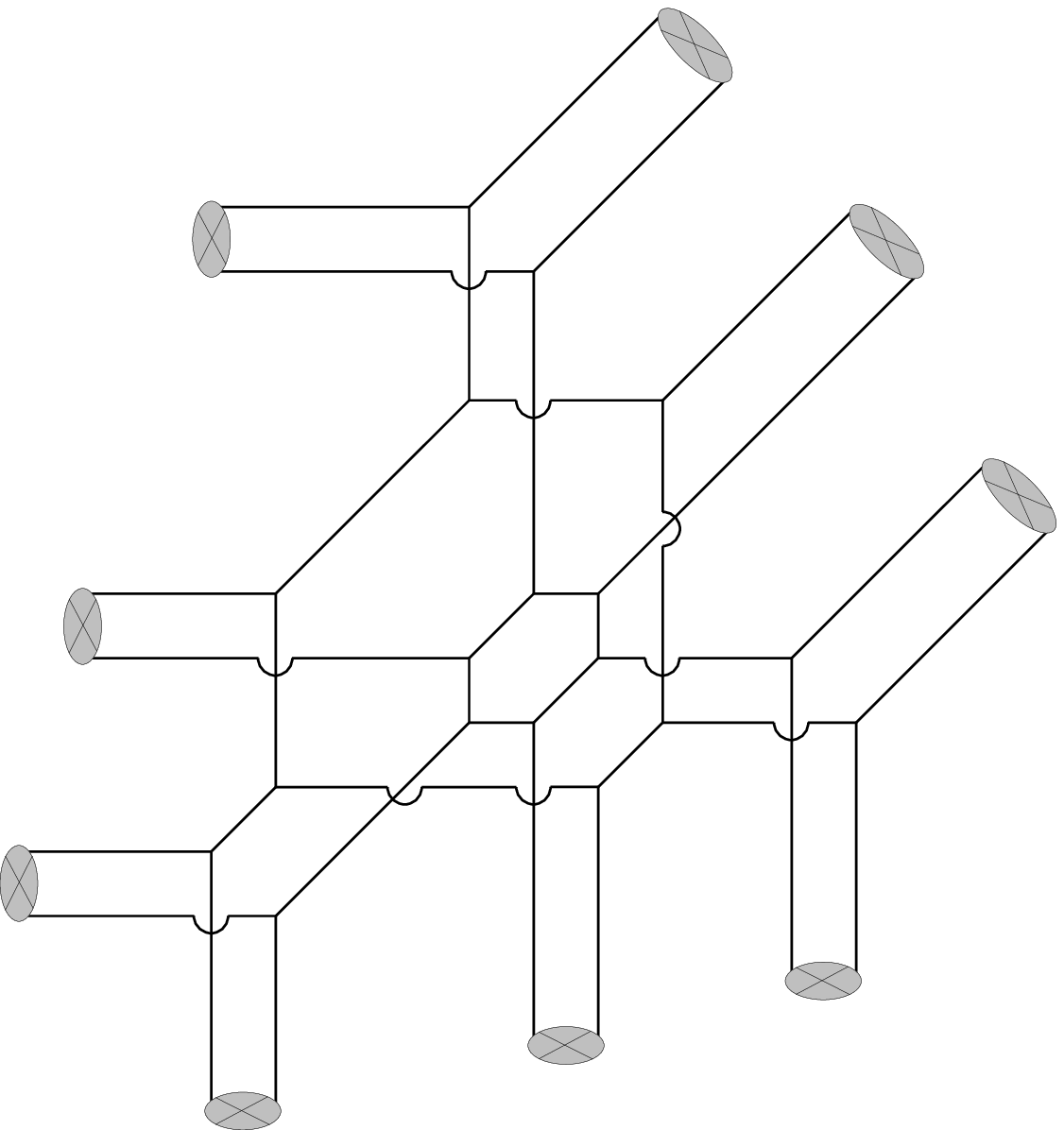}
\hspace{\stretch{1}}
}
\caption{\small The rank $N$ $E_6$ theory, in the example rank $= 2$. On the left: dot diagram with three $\{N^3\}$ partitions on the external edges, tessellated respecting the s-rule. In red the two polygons which become closed faces in the web of 5-branes. On the right: dual web of 5-branes. The two concentric closed faces are well visible.
\label{fig: E6 rank N}}
\end{figure}

There are higher-rank versions of the theories we saw above which have
$E_{6,7,8}$ flavor symmetry, and whose properties are summarized in Appendix~\ref{sec: exp theories}.
We can construct all these theories with the multi-junctions.
To get the rank-$N$ $E_6$ theory, we start from $T[A_{3N-1}]$,
and move on the Higgs branch such that $N$ 5-branes end on the same 7-brane on each side of the multi-junction; therefore  three 7-branes on each edge are needed.
The dimension of the Higgs branch is
\be
\dim_\bbH \calM_H = 3 \cdot 3N + 3N - 1 = 12N - 1 \;.
\ee
We can work out the web of 5-branes that respects the s-rule and count the dimension of the Coulomb branch: $\dim_\bbC \calM_C = N$.
The web of 5-branes turns out to be, due to the jumps,
a superposition of $N$ copies of the $E_6$ web, see Figure \ref{fig: E6 rank N}.

In the same way, we can get the rank $N$ $E_7$ and $E_8$ theories.
More generally, given some set of punctures in a $T[A_{k-1}]$ theory, we can construct a new theory with the same global symmetry but larger Coulomb branch starting with the $T[A_{Nk-1}]$ theory and substituting each 5-brane with $N$ 5-branes ending on the same 7-brane.
More precisely, the number and type of 7-branes in the new theory is the same as in the original one, such that the flavor symmetry is the same. However whenever $m$ 5-branes end on the same 7-brane in the original theory, $Nm$ 5-branes end on it in the new theory. The total number of external 5-branes was $3k$ in the original theory, and is $3Nk$ in the new one. The Higgs branch dimension is easily determined:
\be
\label{N-ification Higgs}
\dim_\bbH \calM_H^\mr{new} +1 = N \big( \dim_\bbH \calM_H^\mr{old} + 1\big) \;.
\ee
The Coulomb branch dimension has to be worked out by tessellating the dot diagram according to the s-rule, and then counting the number of closed faces in the dual web of 5-branes. It turns out that when the dimension  is 1 in the original theory,
the dimension is $N$ in the new theory.

\begin{figure}
\centerline{
\hspace{\stretch{1}}
\includegraphics[height=30ex]{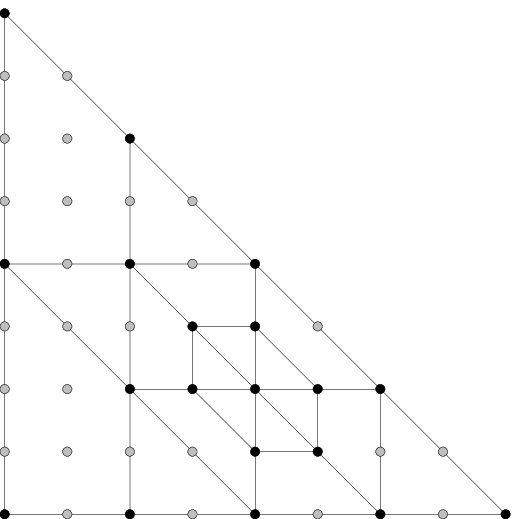}
\hspace{\stretch{2}}
\includegraphics[height=33ex]{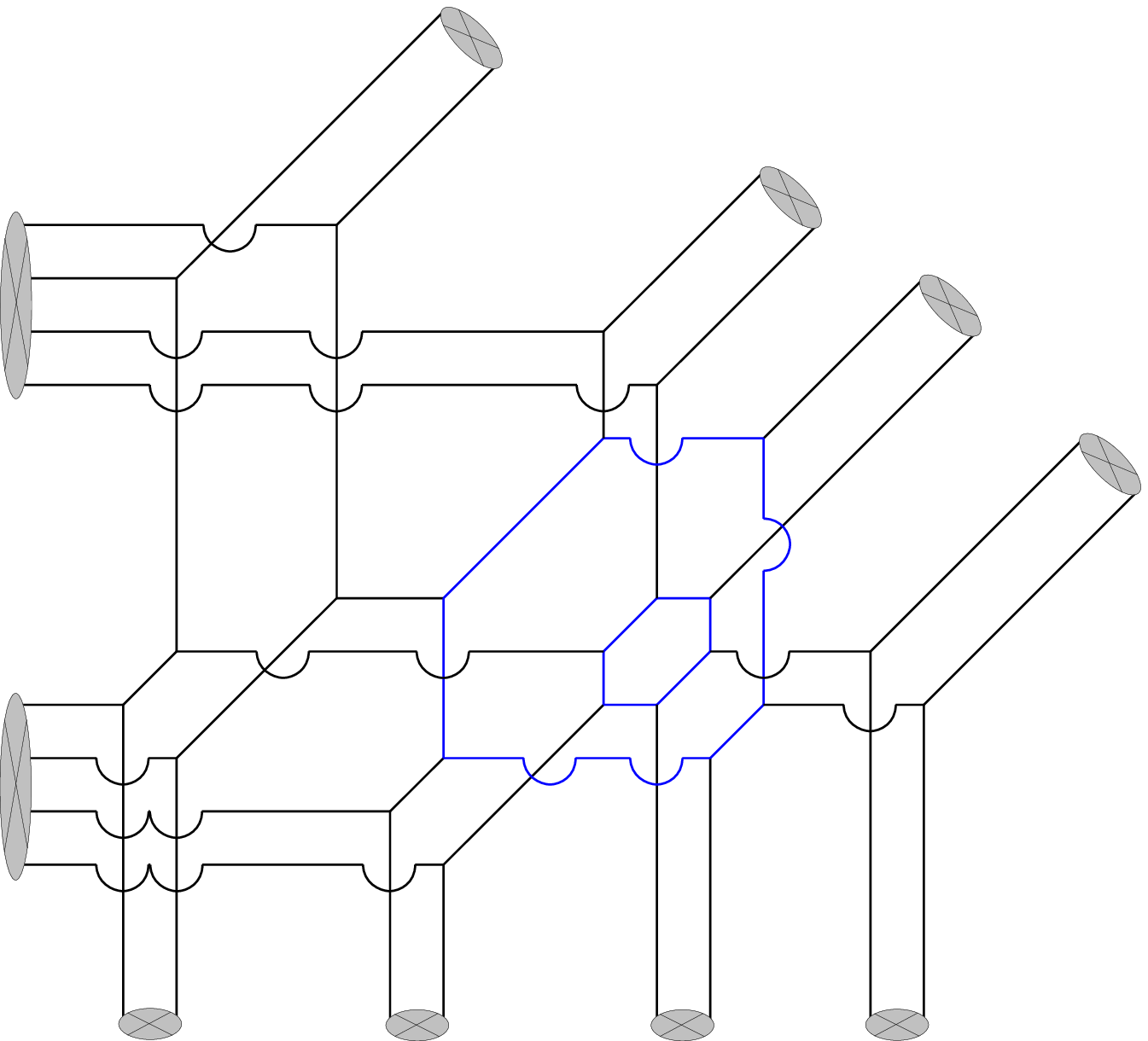}
\hspace{\stretch{1}}
}
\caption{\small The rank $N$ $E_7$ theory, in the example rank $= 2$.  On the left: dot diagram with partitions $\{N^4\}$ and $\{2N,2N\}$ on the external edges. On the right: dual web of 5-branes. The two concentric closed faces are emphasized in blue.
\label{fig: E7 rank N}}
\end{figure}

\begin{figure}
\centerline{
\hspace{\stretch{1}}
\includegraphics[height=30ex]{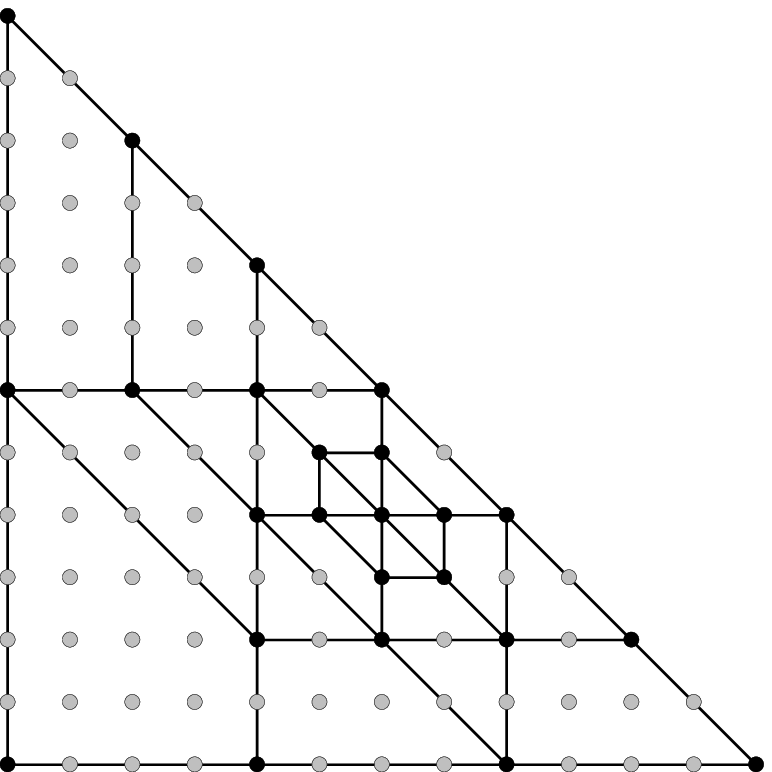}
\hspace{\stretch{2}}
\includegraphics[height=43ex]{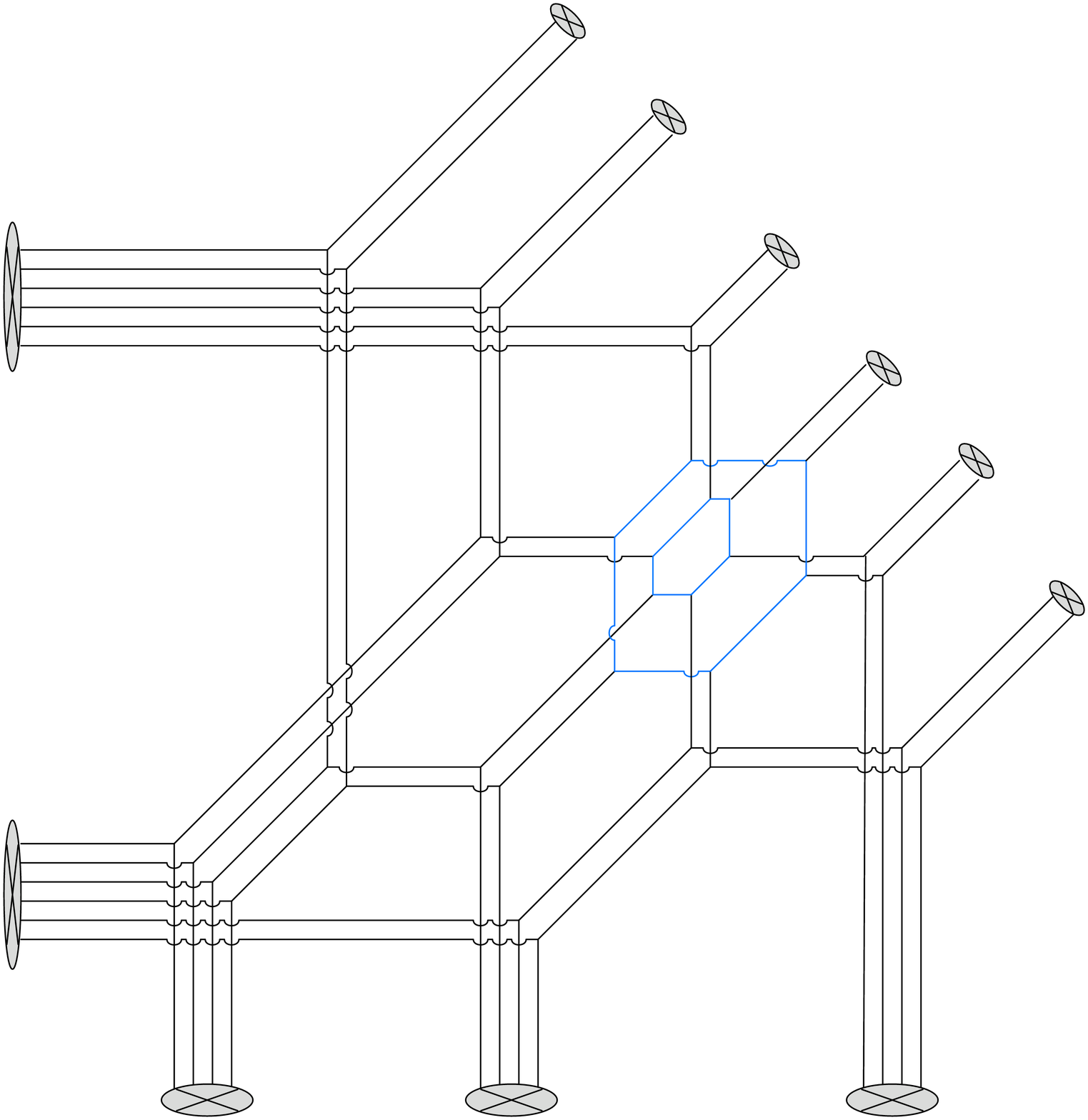}
\hspace{\stretch{1}}
}
\caption{\small The rank $N$ $E_8$ theory, in the example rank $= 2$. On the left: dot diagram with partitions $\{N^6\}$, $\{2N,2N,2N\}$ and $\{3N,3N\}$ on the external edges. On the right: dual web of 5-branes. The two concentric closed faces are emphasized in blue.
\label{fig: E8 rank N}}
\end{figure}

The rank-$N$ $E_7$ theory is embedded in $T[A_{4N-1}]$;
the punctures are two $\{N^4\}$ and one $\{2N,2N\}$.
The Coulomb branch has dimension $N$, see Figure \ref{fig: E7 rank N}, and the Higgs branch dimension $18N-1$.
The rank-$N$ $E_8$ theory is embedded in $T[A_{6N-1}]$.
the punctures are one $\{N^6\}$, one $\{2N,2N,2N\}$ and one $\{3N,3N\}$.
The Coulomb branch has dimension $N$, see Figure \ref{fig: E8 rank N}, and the Higgs branch dimension $30N-1$.

The process can be applied to any multi-junction configuration with three generic punctures. In particular it can be applied to the basic $k$-junction itself, that corresponds to the $T[A_{k-1}]$ SCFT, to obtain a higher rank version of it. The dimensions of the moduli space are
\bea
\dim \calM_H &= \frac{N(3k^2-k)}{2}-1 \\
\dim \calM_C &= \frac{k \big[ (k-9)(N-1) + (k-3)N^2 \big] + 20N - 18}{2} \qquad\text{ for }\qquad k>2 \;.
\eea
The Higgs branch dimension is directly obtained from the general formula (\ref{N-ification Higgs}) and the dimension for the $T[A_{k-1}]$ theory (\ref{Higgs branch max punct}). The Coulomb branch dimension has to be worked out from the dot diagram and the web of 5-branes. We do not know an F-theory construction for these theories.

\section{S-dualities and theories with $E_{6,7,8}$ flavor symmetry}
\label{sec: davide}

In the previous sections, we found that the 4d SCFTs with $E_{6,7,8}$ flavor symmetry
originally found by Minahan and Nemeschansky
can be constructed by means of 5-brane junctions compactified on $S^1$.
Equivalently, they correspond to compactifications of the 6d $A_{N-1}$ theory
on a sphere with three specific punctures, see Table~\ref{table: E}.
\begin{table}
\[
\begin{array}{c|c|ccc|c}
E_n & N & \multicolumn{3}{c|}{\text{punctures}}  & \text{manifest flavor symmetry} \\
\hline
E_6 & 3 & \{1^3\} &\{1^3\} &\{1^3\} & \SU(3)^3 \\
E_7 & 4 & \{1^4\} &\{1^4\} &\{2^2\} & \SU(4)^2\times \SU(2) \\
E_8 & 6 & \{1^6\} &\{2^3\} &\{3^2\} & \SU(6)\times \SU(3)\times \SU(2)
\end{array}
\]
\caption{SCFTs with $E_{6,7,8}$ symmetry via compactifications
of the 6d $A_{N-1}$ theory. For each $E_n$, the number $N$ of M5-branes,
the types of punctures, and the manifest flavor symmetry are listed.
\label{table: E}}
\end{table}
In this section we perform further checks of our proposal using
the formalism by Gaiotto \cite{Gaiotto:2009we}.

\subsection{Formalism}
\label{sec: review}

Let us start by briefly recalling the formalism.
Consider $\cN=2$ superconformal linear quiver gauge theories
with a chain of  $\SU$ groups
\begin{equation}
\label{tac:generallinear}
\SU(d_1)\times \SU(d_2) \times\cdots \times \SU(d_{n-1}) \times \SU(d_n) \;,
\end{equation}
a bifundamental hypermultiplet between each pair of consecutive gauge groups
$\SU(d_a)\times \SU(d_{a+1})$,
and $k_a$ extra fundamental hypermultiplets for $\SU(d_a)$.
We require $k_a= 2d_a - d_{a-1} - d_{a+1}$
to make every gauge coupling marginal, where we defined $d_0=d_{n-1}=0$.
Since $k_a$ is non-negative, we have
\begin{equation}
d_1 < d_2 <  \cdots  < d_l  = \cdots = d_r > d_{r+1} > \cdots > d_n \;.
\end{equation}
We denote $N=d_l=\cdots =d_r$;
we refer to the parts to the right of $d_r$ and to the left of $d_l$
as  two tails of this superconformal quiver.
Requirement that $k_a\ge 0$ means that $d_a-d_{a+1}$ is monotonically non-decreasing for $a>r$;
thus we can associate naturally a Young tableau to the tail
by requiring that it has a row of width $d_a-d_{a+1}$ for each $a\ge r$.
Therefore we can naturally label a puncture by a Young tableau.

The Seiberg-Witten curves for these quivers were originally found in \cite{Witten:1997sc}.
It was then shown  in  \cite{Gaiotto:2009we} that they can be realized as a subspace
of the total bundle $T^*\Sigma$
of holomorphic differentials on a Riemann surface $\Sigma$,
given by the equation
 \begin{equation}
0=x^N+x^{N-2}\phi_2 + x^{N-3}\phi_3+ \cdots+ \phi_N \;,
\end{equation}
where $x$ is a holomorphic differential on the Riemann surface $\Sigma$ which parameterizes the fiber direction,
and $\phi_d$ is a degree-$d$ differential with poles at the punctures,
encoding VEV's of Coulomb branch operators of dimension $d$.
We call this set of a Riemann surface $\Sigma$ and punctures marked by
Young tableaux the {\em G-curve} of the system, to distinguish it from the Seiberg-Witten curve.
One finds that, for the general quivers \eqref{tac:generallinear},
one has $n+1$ punctures of the same type, which we call `simple punctures' and denote by $\bullet$,
and two extra punctures labeled by Young tableaux
which  encode the information on  the tails.

At a simple puncture $\phi_d$ is allowed to have a simple pole.
At a more general puncture, $\phi_d$ is allowed to have a pole of higher order.
We denote by $p_d$ the order of the pole which $\phi_d$ is allowed to have at the puncture;
the method to obtain $p_d$ from the Young tableau was detailed in \cite{Gaiotto:2009we}. At a given puncture specified by the partition $\{k_i\}$ and the corresponding Young tableau, the orders $p_d$ of the allowed poles of the degree-$d$ differentials $\phi_d$ are determined as follows. The Young tableau has columns of height $k_J \geq \dots \geq k_1$, aligned at the bottom. We order the $N$ boxes from left to right and then from bottom up, starting from the bottom left corner. Then $p_i = i - h(i)$, where $h(i)$ is the height of the $i$-th box, and the bottom row has height 1.
The number of the Coulomb branch operators of dimension $d$
is then given by the dimension of the space of degree-$d$ differentials
with the prescribed singularities. The formula on the sphere is
\begin{equation}
\label{tac:formula}
\#\ \text{of operators of dim. $d$ }
=\sum_\text{punctures} (\text{$p_d$ at the puncture}) - (2d-1) \;.
\end{equation}

The marginal couplings of the theory are encoded in
the shape of the punctured Riemann surface $\Sigma$.
For example, by studying the quivers of the form \eqref{tac:generallinear}
we can show that  a sphere containing $N-1$ simple punctures splits off
and leaves a puncture labeled by the tableau with one row of $N$ boxes,
when the coupling of the $\SU(N-1)$ group
inside a superconformal tail with the gauge groups \begin{equation}
\SU(N-1)\times\SU(N-2)\times \cdots \times \SU(2)
\end{equation} becomes weak.
Now, this splitting of $N-1$ simple punctures can also occur
in the very strongly-coupled regime. Following \cite{Argyres:2007cn,Gaiotto:2009we}
we identify this situation as having a weakly-coupled S-dual description,
with the superconformal tail of the form above arising non-perturbatively.

\subsection{Rank-$1$ $E_n$ theories}

\begin{figure}
\[
\begin{array}{c@{\qquad}c}
\inc{.3}{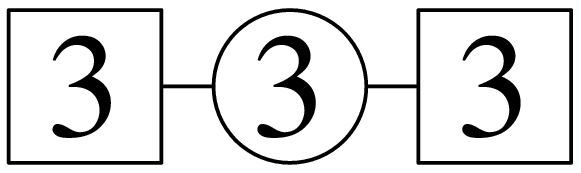} & \inc{.3}{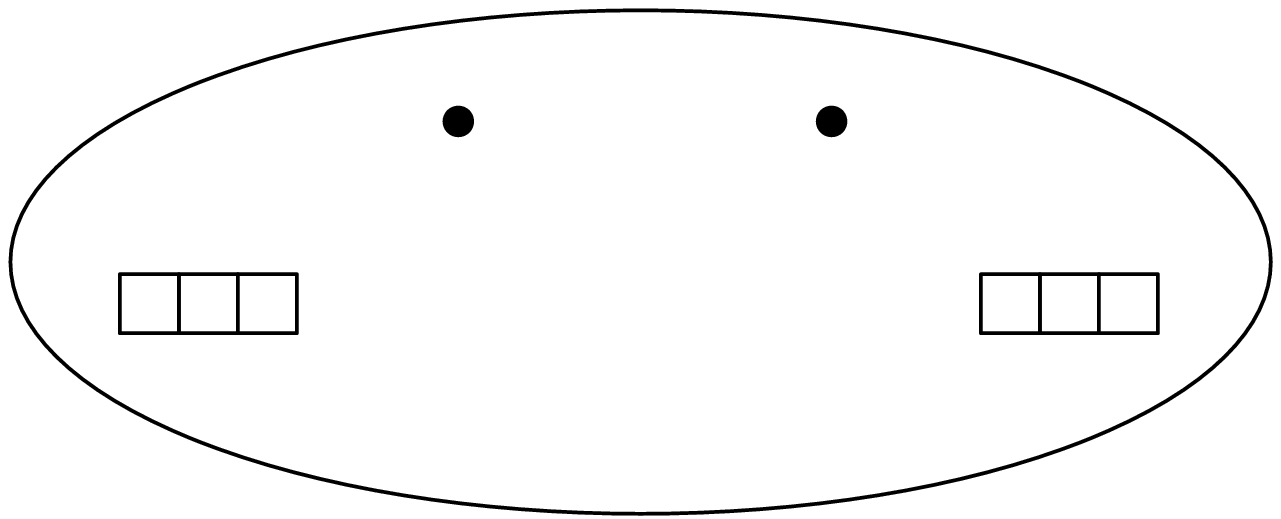} \\
\Downarrow & \Downarrow \\
\inc{.3}{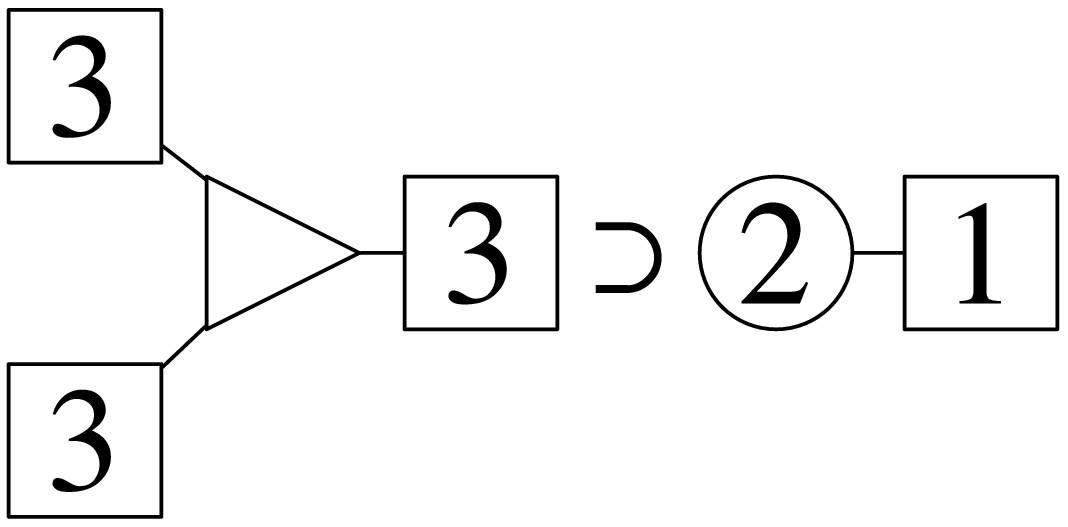} & \inc{.3}{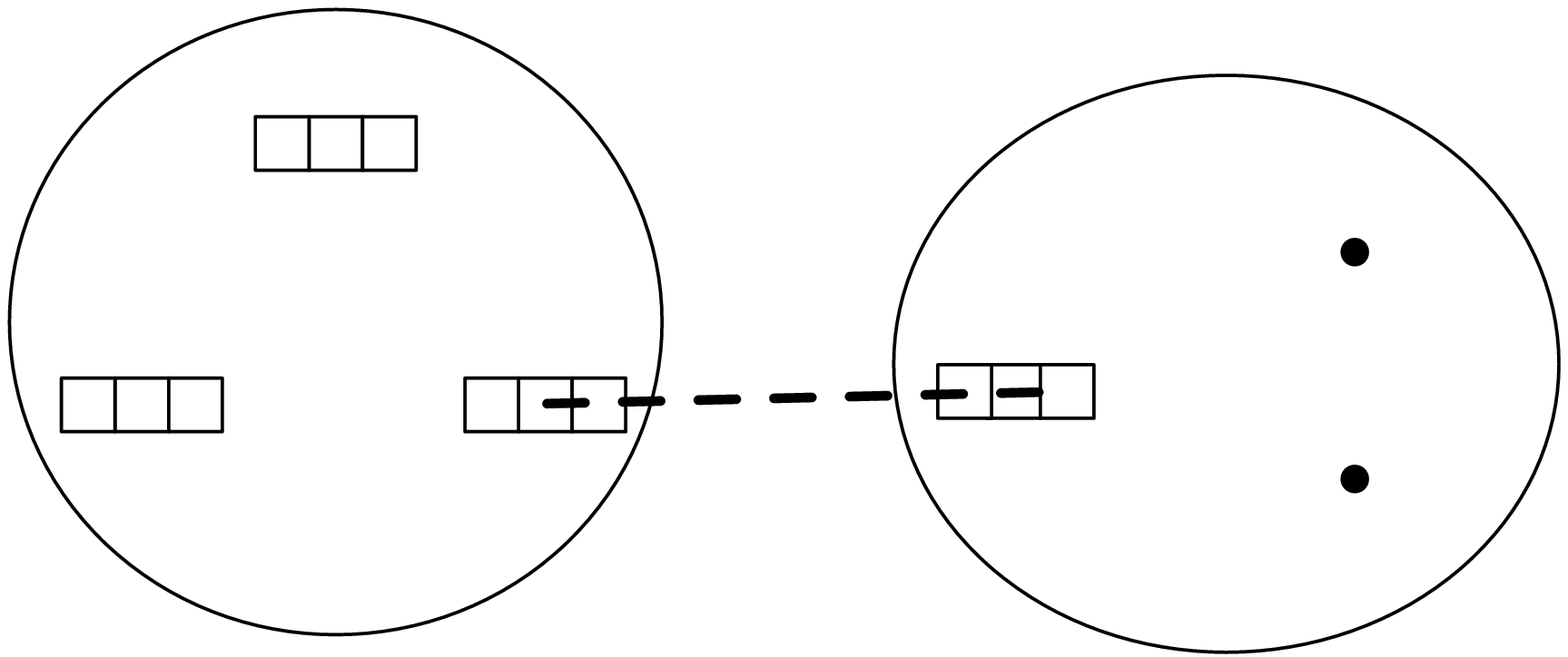}
\end{array}
\]
\caption{Construction of the rank-1 $E_6$ theory.
A circle or a box with a number $n$ stands for a $\SU(n)$ gauge group or flavor symmetry, respectively. The line connecting two objects stands for a bifundamental hypermultiplet charged under two groups. The symbol $\supset$ between a flavor symmetry and a gauge symmetry signifies that the gauge fields coupled to the subgroup of the flavor symmetry specified. The triangle with three $\SU(3)$ flavor symmetries is the Minahan-Nemeschansky's $E_6$ SCFT. The G-curve is shown on the right. \label{tac:E6}}
\end{figure}

The way to obtain the rank-1 $E_6$ theory as a limit of a field theory
with Lagrangian was first obtained in \cite{Argyres:2007cn};
we follow the presentation in \cite{Gaiotto:2009we}.
The construction starts from the quiver shown in the first line of Fig.~\ref{tac:E6},
whose G-curve is also shown there. It is an $\SU(3)$ gauge theory with six
hypermultiplets in the fundamental representation.
The limit where the coupling constant of $\SU(3)$ is infinitely strong
corresponds to the degeneration of the G-curve such that two simple punctures
of type $\bullet$ come together and develop a neck.
A dual weakly-coupled $\SU(2)$ gauge group with one flavor appear.
In the zero coupling limit of this new $\SU(2)$ gauge group, the neck pinches off
and produces another puncture $\{1^3\}$.
We end up with a theory whose G-curve is a sphere with three punctures of type $\{1^3\}$.
On the one hand, in the original description as an $\SU(3)$ gauge theory with six flavors,
it was manifest that $\SU(3)^2$ enhances to $\SU(6)$.
On the other hand, in the description with the G-curve, it is manifest that
the three $\SU(3)$ flavor symmetries are on the same footing; therefore
any pair of two out of the three $\SU(3)$ groups should enhance to $\SU(6)$,
which is possible only if this theory has $E_6$ flavor symmetry.%
\footnote{The authors thank Davide Gaiotto for explaining this argument of the enhancement to $E_6$.}

\begin{figure}
\[
\begin{array}{c@{\qquad}c}
\inc{.3}{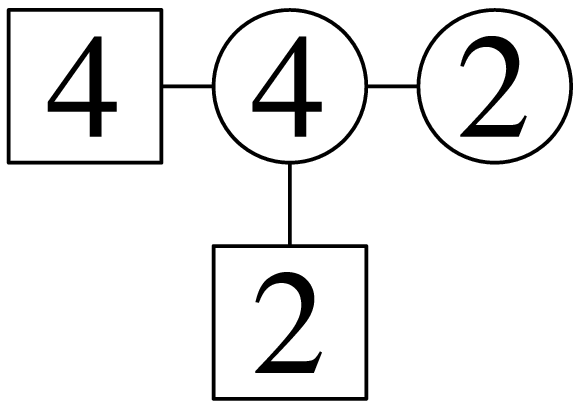} & \inc{.3}{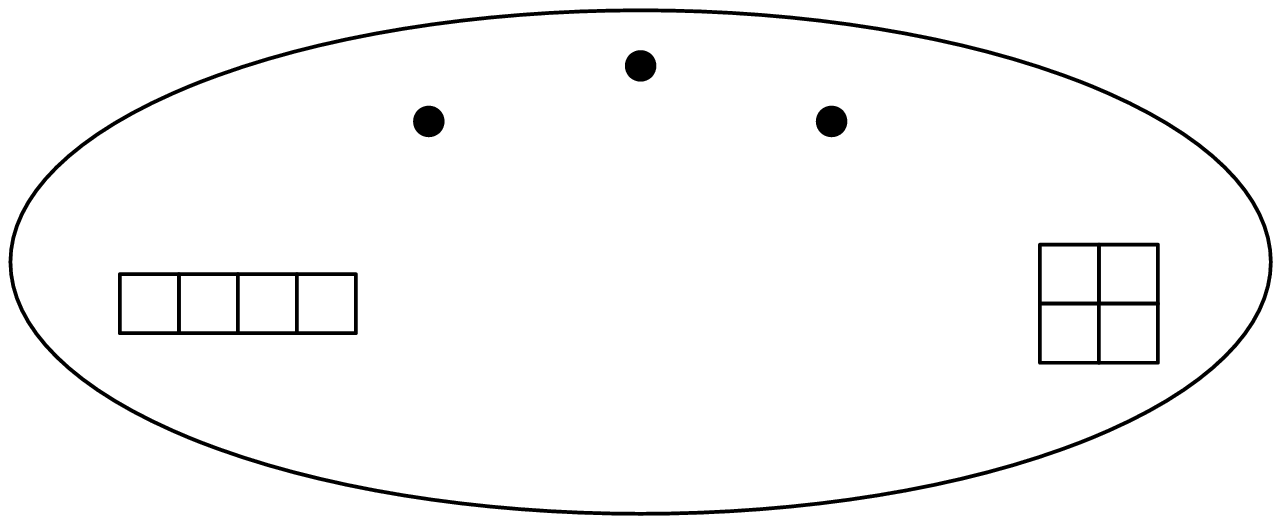} \\
\Downarrow & \Downarrow \\
\inc{.3}{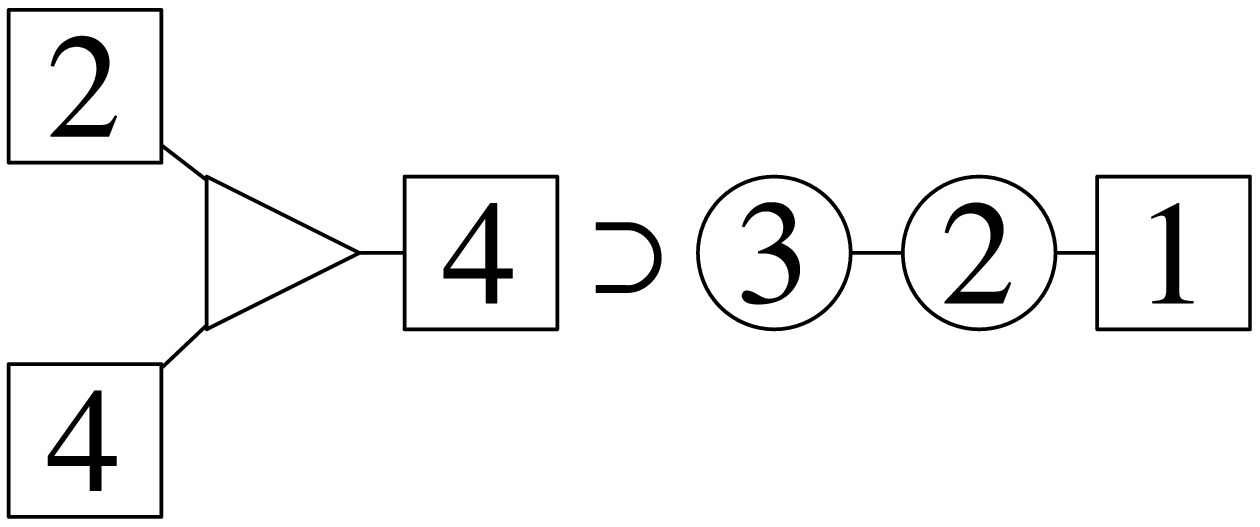} & \inc{.3}{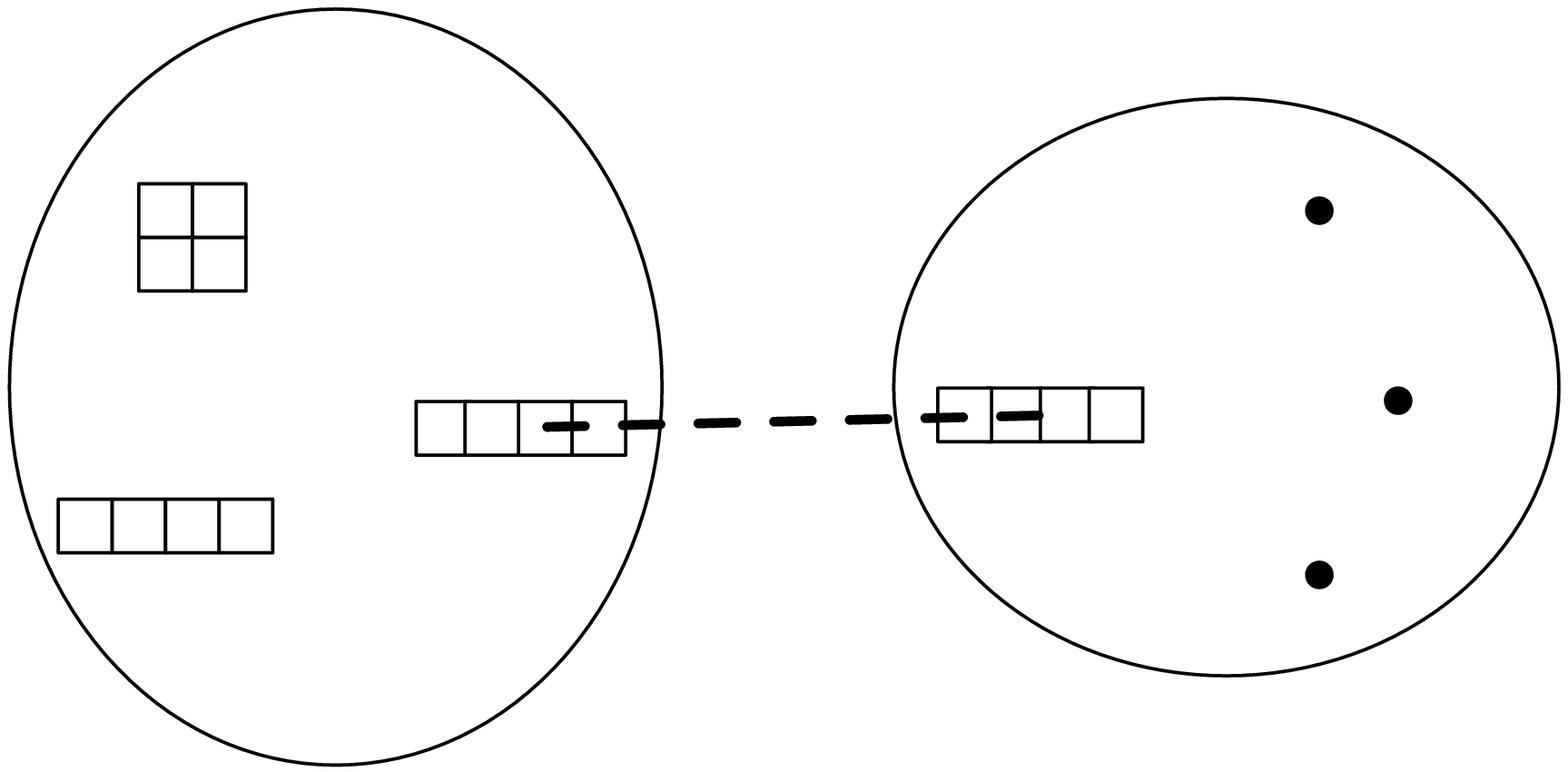}
\end{array}
\]
\caption{Construction of the rank-1 $E_7$ theory.
 The triangle with two $\SU(4)$ and one $\SU(2)$ flavor symmetries represents the Minahan-Nemeschansky's $E_7$ SCFT.\label{tac:E7}}
\end{figure}

The $E_7$ theory was found in the infinitely strongly-coupled limit
of a $\USp(4)$ gauge theory with six fundamental hypermultiplets
in \cite{Argyres:2007cn},
which was also directly realized in the quiver language in \cite{Gaiotto:2009we}.
Another realization was recently found in \cite{Tachikawa:2009rb}.
Here instead, we present a method to construct it using a quiver consisting
solely of $\SU$ groups.
We start from the quiver shown in the first line of Fig.~\ref{tac:E7}.
The gauge group is $\SU(4)\times \SU(2)$ with the bifundamental
hypermultiplets charged under the two $\SU$ factors,
and there are in addition six fundamental hypermultiplets for the node $\SU(4)$.
Its G-curve has three simple punctures,
one puncture $\{1^4\}$ and one $\{2^2\}$.
We can go to a limit where a sphere with three simple punctures splits.
A dual superconformal tail with gauge groups $\SU(3)\times \SU(2)$ appears.
After the neck is pinched off,
we have a theory whose G-curve is a sphere with one puncture of type $\{2^2\}$
and two punctures of type $\{1^4\}$.
This description shows the flavor symmetry $\SU(2)\times \SU(4)^2$.
In the original description, it is clear that $\SU(2)\times \SU(4)$ enhances to
$\SU(6)$. In the description using the G-curve,
the two $\SU(4)$ cannot be distinguished. This is only possible
when the total flavor symmetry enhances to $E_7$.

\begin{figure}
\[
\begin{array}{c@{\qquad}c}
\inc{.3}{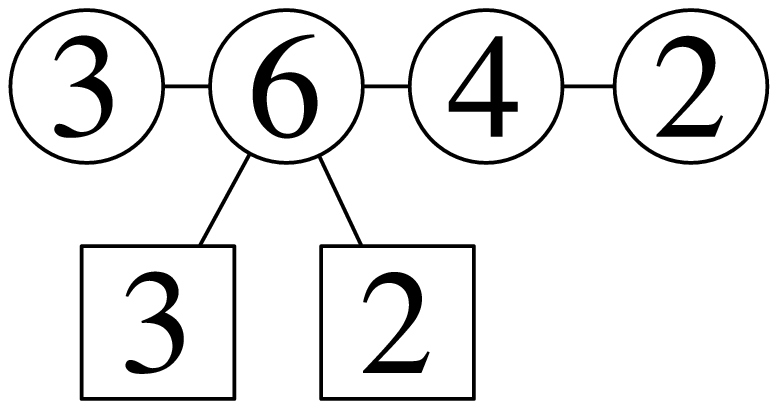} & \inc{.3}{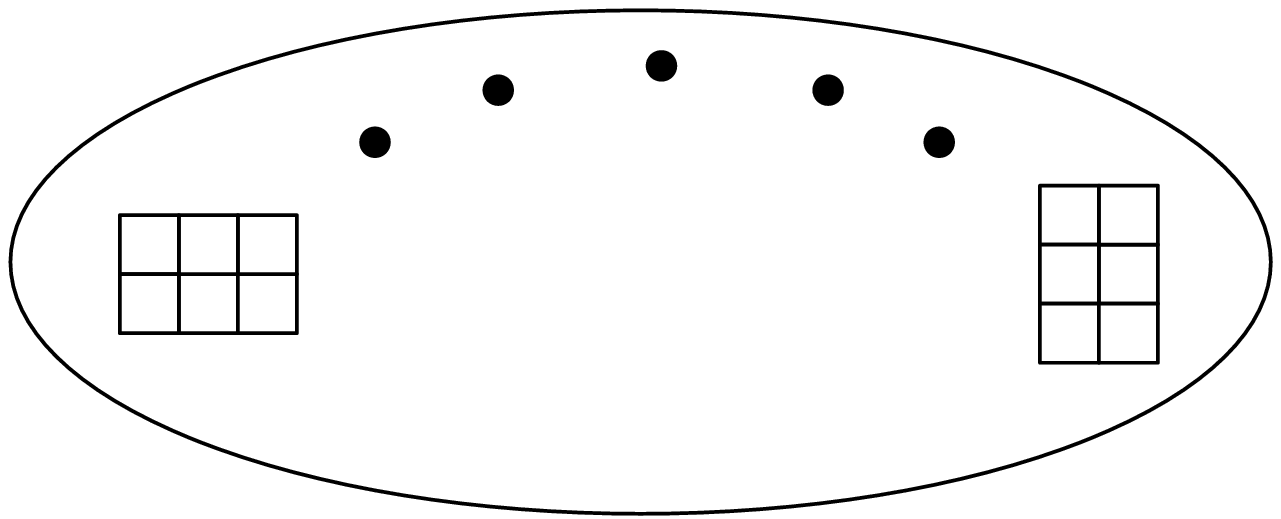} \\
\Downarrow & \Downarrow \\
\inc{.3}{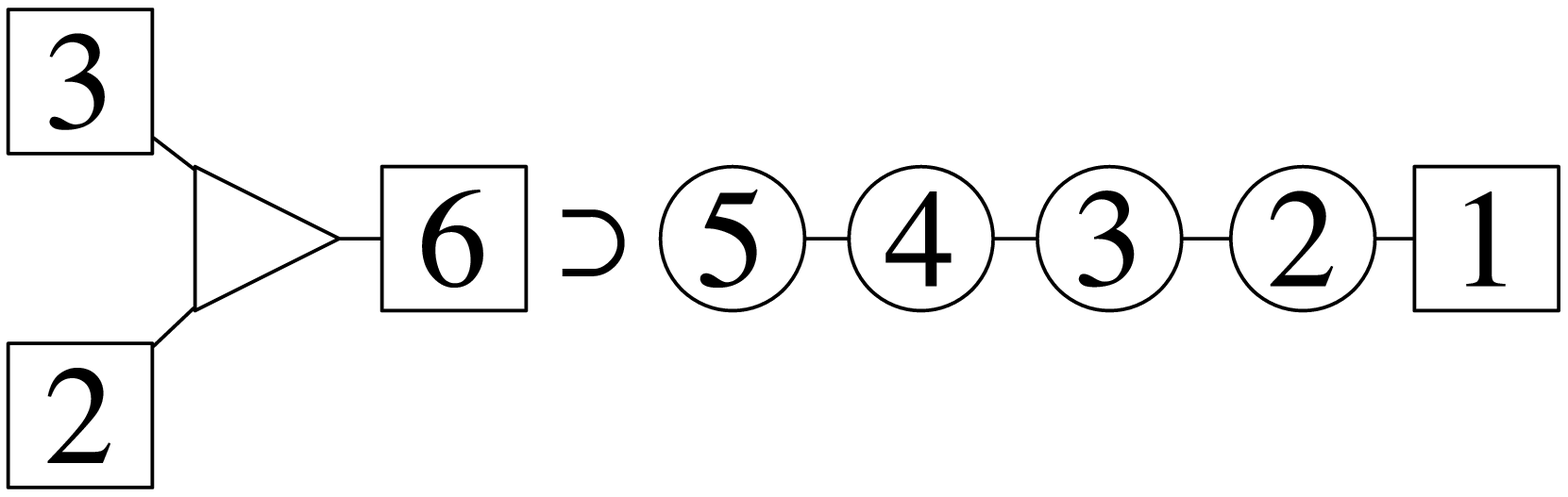} & \inc{.3}{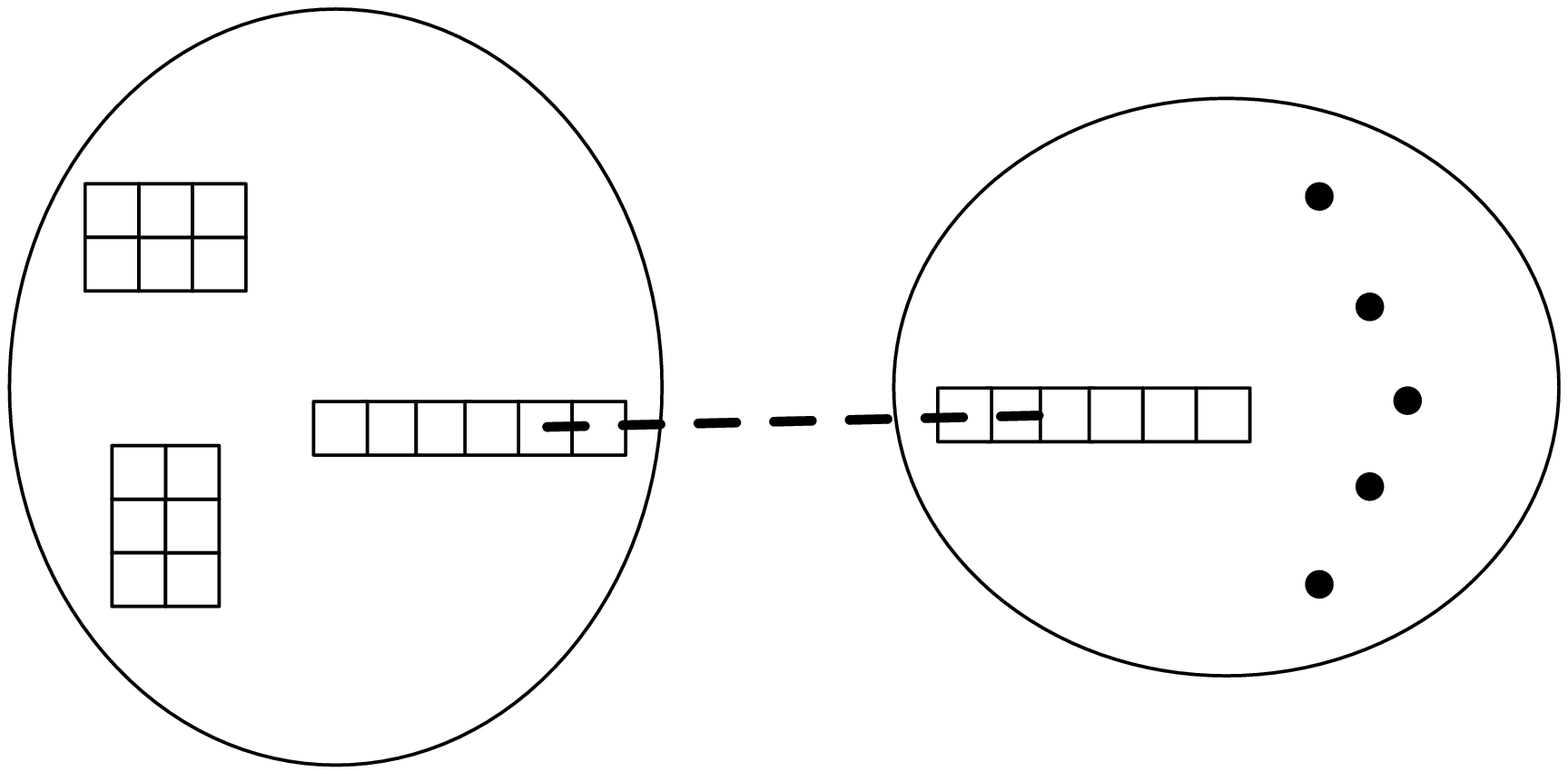}
\end{array}
\]
\caption{Construction of the rank-1 $E_8$ theory.
 The triangle with one $\SU(6)$, one $\SU(3)$  and one $\SU(2)$ flavor symmetries stands for the Minahan-Nemeschansky's $E_8$ SCFT.\label{tac:E8}}
\end{figure}

The rank-1 $E_8$ theory was found in the infinitely strongly-coupled limit
of several kinds of Lagrangian field theories in \cite{Argyres:2007tq};
the realization we present here does not seem to be directly related
to the cases listed there.
We start from the quiver shown in the first line of Fig.~\ref{tac:E8}.
The original quiver has the gauge group \begin{equation}
\SU(3)\times \SU(6)\times \SU(4)\times \SU(2)
\end{equation} with bifundamental hypermultiplets between the consecutive
$\SU$ factors; one has in addition five fundamental hypermultiplets for $\SU(6)$.
Its G-curve has five simple punctures,
one puncture $\{2^3\}$ and one $\{3^2\}$.
We can go to a limit where a sphere with five simple punctures splits off.
A dual superconformal tail with gauge groups
\begin{equation}
\SU(5)\times \SU(4)\times \SU(3)\times \SU(2)
\end{equation}
appears. We tune the gauge coupling of the $\SU(5)$ group to zero,
leaving a theory whose G-curve is a sphere with one puncture of type $\{1^6\}$,
one of type $\{2^3\}$ and another of type $\{3^2\}$.
This description shows the flavor symmetry $\SU(2)\times \SU(3)\times \SU(6)$.
In the original description, it is clear that $\SU(3)\times \SU(2)$ enhances to
$\SU(5)$.
This $\SU(5)$ does not commute  with
the $\SU(6)$ associated to the puncture $\{1^6\}$, because if it did,
the generalized quiver drawn in the second line of Fig.~\ref{tac:E8}
would have $\SU(5)\times \U(1)^5$ symmetry while the original quiver clearly has
only $\SU(5)\times \U(1)^4$.
The only possibility is that $SU(5)$ and $SU(6)$ combine to form $E_8$ (we refer the reader to Tables 14 and 15 in \cite{Slansky:1981yr}).

Indeed, the structure of the Coulomb branch indicates that this is the $E_8$ theory
of Minahan and Nemeschansky.
It can be easily found, using the formula \eqref{tac:formula}, that the theory whose G-curve has three punctures of type
$\{1^6\}$,  $\{2^3\}$,  $\{3^2\}$ has only one Coulomb branch operator,
whose dimension is 6. This agrees with the known fact
of the $E_8$ theory.
One can also easily calculate the central charges $a$ and $c$,
or equivalently the effective number $n_v$ and $n_h$ of hyper- and vector multiplets.
In the original linear quiver, we have
\begin{equation}
n_v(\text{total})=61 \;, \qquad\qquad n_h(\text{total})=80 \;,
\end{equation}
whereas the tail contains
\begin{equation}
n_v(\text{tail})=50 \;, \qquad\qquad n_h(\text{tail})=40 \;.
\end{equation}
We conclude that
\begin{equation}
n_v(E_8)=11 \;, \qquad\qquad n_h(E_8)=40
\end{equation}
or equivalently
\begin{equation}
a(E_8)=\frac{95}{24} \;, \qquad\qquad c(E_8)=\frac{31}{6} \;.
\end{equation}
They agree with what was found in \cite{Aharony:2007dj,Argyres:2007tq}.

The same procedure
works for the $E_6$  and $E_7$ theories treated above.
The result is that they have only one Coulomb branch operator each,
of dimension $3$ and $4$ respectively, which again agrees with the known properties
of these theories. The central charges $a$ and $c$ can also be easily calculated,
correctly reproducing the known data.

\subsection{Higher-rank $E_n$ theories}

In the previous section we argued that
the theory with three punctures of type $\{N^3\}$
has the right properties to be identified with the higher-rank $E_6$ theory.
In this section we provide further pieces of evidence.
First, let us determine the spectrum of Coulomb branch operators.
Using the algorithm explained in Sec.~\ref{sec: review},
the poles of the degree-$d$ differential $\phi_d$  at the puncture of type $\{N^3\}$
have degrees
\begin{equation}
(p_2,p_3;\, p_4,p_5,p_6;\, p_7,p_8,p_9;\, \ldots,p_{3N})
= (1,2;\, 2,3,4;\, 4,5,6;\, \ldots, 2N) \;.
\end{equation}
Using \eqref{tac:formula}, we find that  this theory has operators of dimension
\begin{equation}
3,\, 6,\, 9,\, \ldots, 3N
\end{equation}
and the number of operators of each dimension is one.
This agrees with the known fact of the rank-$N$ $E_6$ theory.

\begin{figure}
\[
\inc{.3}{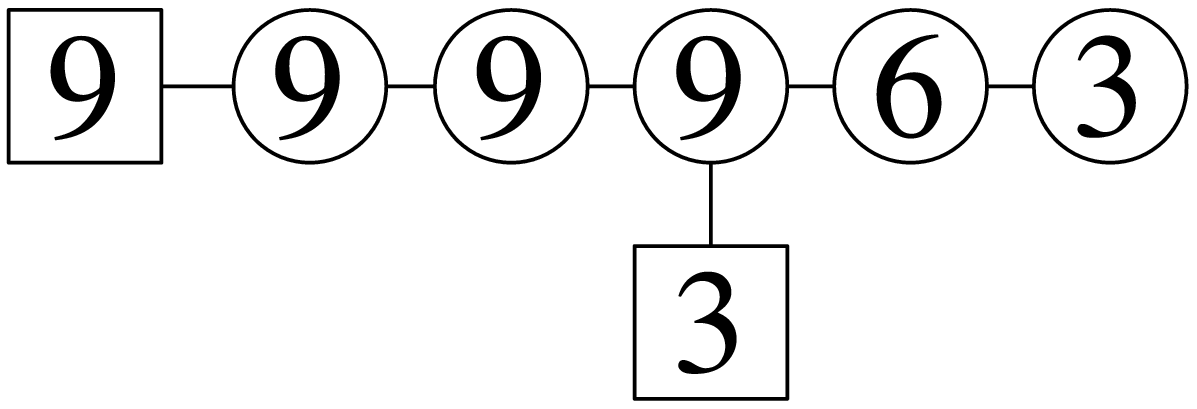} \qquad
\inc{.3}{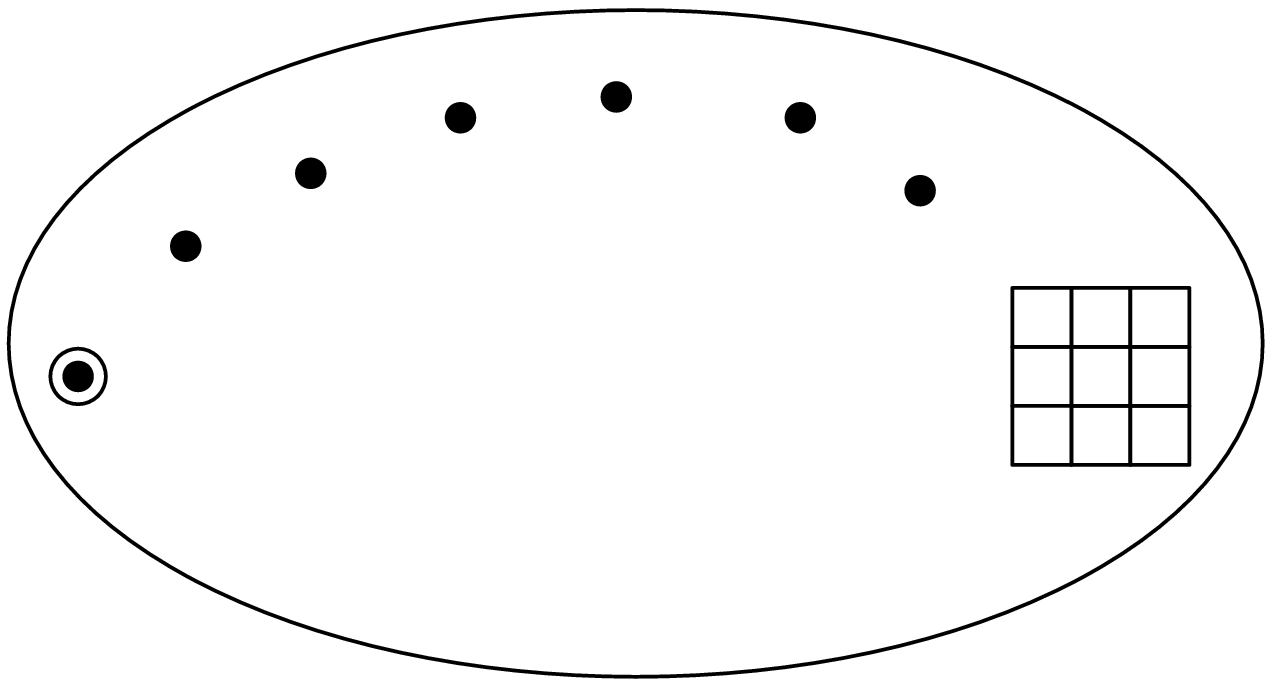}
\]
\caption{A linear quiver whose G-curve has a tableau $\{N^3\}$. Here $N=3$. \label{tac:example}}
\end{figure}

The central charge of the $\SU(3)$ flavor symmetry can also be determined;
the tableau $\{N^3\}$ appears for instance in the G-curve of a quiver
with gauge groups
\begin{equation}
\SU(3N)^a \times \SU(3N-3)\times \SU(3N-6)\times \SU(6)\times \SU(3) \;,
\end{equation}
with $3N$ and $3$ hypermultiplets in the fundamental representation
for the leftmost and the rightmost $\SU(3N)$ gauge groups to make them superconformal.
See Fig.~\ref{tac:example} for the case $a=3$, $N=3$.
There are $3N$ hypermultiplets transforming under the $\SU(3)$ flavor symmetry,
therefore we have
\begin{equation}
k_{\SU(3)}= 6N
\end{equation}
which is consistent with the known fact $k_{E_6}=6N$ for the
rank-$N$ $E_6$ theory.

We would also like to compute the central charges $a$ and $c$
of the superconformal current,
or equivalently the effective numbers of vector and hypermultiplets $n_v$ and $n_h$
of this theory, and compare them to the known values.
Unfortunately it is not known how to construct this theory
in the framework of \cite{Gaiotto:2009we}, in which the class of theories
of this type was called `unconstructible'. Our web construction suggests that these
theories can be found along the Higgs branch of a parent theory, rather then
at corners of its marginal coupling parameter space. However this procedure involves
an RG flow, along which $a$ and $c$ generically vary.
It would be worthwhile to study such theories further, and to determine their central charges.

The analysis for our candidate higher-rank $E_{7,8}$ theories are similar.
For the $E_7$ theory, the candidate is a theory whose G-curve has two punctures of
type $\{N^4\}$ and one of type $\{2N,2N\}$. The pole structure at the puncture $\{N^4\}$ is
\be
(p_2, p_3, p_4;\, p_5, p_6, p_7, p_8;\, \dots p_{4N}) = (1,2,3;\, 3,4,5,6;\, \dots 3N) \;,
\ee
while at the puncture $\{2N,2N\}$ is
\be
(p_2;\, p_3, p_4;\, p_5, p_6;\, p_7, p_8;\, \dots p_{4N}) = (1;\, 1,2;\, 2,3;\, 3,4;\, \dots 2N) \;.
\ee
Combined with the formula \eqref{tac:formula} above, one concludes that
this theory has Coulomb branch operators of dimension
\begin{equation}
4,\, 8,\, \ldots, 4N \;,
\end{equation}
each with multiplicity one.
The manifest flavor symmetry is $\SU(2)\times\SU(4)^2$,
and the flavor symmetry central charges are easily found to be
\begin{equation}
k_{\SU(2)}=k_{\SU(4)} = 8N \;,
\end{equation}
which is consistent with the known result $k_{E_7}=8N$
for the rank-$N$ $E_7$ theory.
The case for $E_8$ is left as an exercise to the reader.

\section{Future directions}
\label{sec: directions}

In this paper we proposed that all the isolated four-dimensional \Nugual{2} SCFT's constructed in \cite{Gaiotto:2009we} by wrapping $N$ M5-branes on a sphere with 3 generic punctures, can be equivalently obtained starting with a web configuration of 5-branes in type IIB string theory suspended between parallel 7-branes, and then further compactifying the resulting low energy 5d field theory on $S^1$. This alternative construction plays the role that systems of D4-branes suspended between NS5-branes and D6-branes play in order to describe linear or elliptic quivers of four-dimensional $SU$ gauge groups with extra fundamental matter, or systems of D3-branes suspended between 5-branes play to describe 3d field theories. In particular, the dimension and structure of the Coulomb and Higgs moduli space as well as mixed branches become manifest.

In the particular case in which all three punctures are maximal, that is each external 5-brane ends on its own 7-brane and thus the type is $\{1^N\}$,
the 7-branes could be actually removed without changing the low energy 5d and consequently 4d dynamics. This allows us to use known string dualities, and to argue that the low energy dynamics of M-theory on the CY$_3$ singularity $\bbC^3/\bbZ_N \times Z_N$ is described by a 5d version of the $T[A_{N-1}]$ theory, while after further compactification on $S^1$ we get the 4d isolated SCFT $T[A_{N-1}]$.  Another chain of dualities leads to write down the SW curve of the compactified 5d theory, and in a suitable scaling limit the SW curve of the 4d theory.

The first interesting question is how can 7-branes be incorporated in this chain of dualities.
For instance, one would like to obtain the SW curve for the theory with three generic punctures. We already know what the result is \cite{Gaiotto:2009we}, however it would be interesting to rederive it from the web construction. In particular, in the duality from type IIB on a circle and M-theory on a torus, we find a
$T^2$ fibration over $\bbR^2$,
with a $[p,q]$ 7-brane mapped to a point where the $(p,q)$ one-cycle
of the torus  shrinks.
This space is essentially the elliptically fibered 2-fold which gives the F-theory description of the 7-brane.
The web of 5-branes is thus mapped to a single M5-brane wrapping an holomorphic curve in the 2-fold, and this curve is exactly the SW curve of the compactified 5d theory. It would be interesting to work out this curve explicitly.

On the other hand, another chain of dualities maps the web to pure geometry in M-theory. When 7-branes are present, the dual geometry is non-toric. For the particular case of the $E_n$ theories, the dual geometry is known to be the total space of the canonical line bundle of the del Pezzo  $dP_n$. It would be interesting to understand other examples, for instance the higher rank $E_{6,7,8}$ theories, or investigating whether the dot diagrams we introduced can be helpful in the study of non-toric geometry.

Another question is the extension of the web construction to $SO$ and $USp$ groups. From the point of view of the compactification of the six-dimensional $(2,0)$ $D_N$ theory on a Riemann surface, the problem was analyzed in \cite{Tachikawa:2009rb}. In our construction, isolated SCFT's with $SO$ and possibly $USp$ global symmetries should arise after the introduction of orientifold planes.

One could also be interested in realizing all SCFT's presented in \cite{Gaiotto:2009we}, that is those with more than three punctures on the sphere, and more generally those arising from the compactification of M5-branes on higher genus surfaces. Even though we have not discussed this in the paper, it is indeed possible to glue together the multi-junctions by attaching two bunches of $N$ semi-infinite 5-branes. They correspond to the `constructible' theories in \cite{Gaiotto:2009we}. However some care is required because the parent 5d theories have generically more rich dynamics at particular points of their parameter space with respect to the pure 4d theories. Such problems do not arise for the multi-junction.

Finally, \Nugual{2} theories have the property that, moving along the Coulomb branch, they can be deformed in such a way that a cascading RG flow takes place in which the ranks of the non-Abelian gauge groups progressively reduce \cite{Polchinski:2000mx, Aharony:2000pp, Benini:2008ir}. It could be interesting to study if similar phenomena take place in the present case.

\section*{Acknowledgments}
We thank Davide Gaiotto for his patience and willingness
to teach us the content of his paper. We also thank Diego Rodr\'\i guez-G\'omez for collaboration at an early stage of this project. F.~B. acknowledges the kind hospitality of the Aspen Center for Physics during the completion of this work.

F.~B. is supported by the US Department of Energy under grant No. DE-FG02-91ER40671.
S.~B. is supported in part by the National Science Foundation under
Grant No. PHY-0756966.
Y.~T. is supported in part by the NSF grant PHY-0503584, and in part by the Marvin L.
Goldberger membership at the Institute for Advanced Study.

\appendix

\section{Seiberg-Witten curves}
\label{sec: SW curve}

The Seiberg-Witten curves for the 5d and 4d low energy theories are readily obtained from our construction \cite{Kol:1997fv, Brandhuber:1997ua, Aharony:1997bh}. If we compactify the 5d theory on a circle $x^4 \simeq x^4 + L_B$, then type IIB on $S^1$ is dual to M-theory on a torus of modular parameter $\tau$ equal to the axiodilaton $\tau_B$ of type IIB. The relation between IIB and M-theory quantities is:
\be
L_t = \frac{ g_{sB} (2\pi)^2 \alpha'}{L_B} \qquad\qquad L_A = \frac{(2\pi)^2 \alpha'}{L_B} \qquad\qquad l_p^3 = \frac{L_t \alpha'}{2\pi} \;,
\ee
where $L_t$ and $L_A \equiv L_t \im\tau$ are the lengths of the two sides of the M-theory torus (the area is $L_t L_A$) and $l_p$ is the 11d Planck length.
In the duality between IIB on $S^1$ and M-theory on $T^2$, the web of 5-branes is mapped to an M5-brane wrapping a curve in $T^2 \times \bbR^2$. The curve is obtained from the toric diagram  through:
\be
\label{5d curve}
0 = F(\alpha,\beta) = \sum_{\text{dots }(i,j)} C_{i,j} \, \alpha^i \beta^j \;,
\ee
where we sum over the dots of the toric diagram, $(i,j)$ are the integer coordinates of the dots, $C_{i,j}$
are parameters, and $(\alpha,\beta) \in \bbC^* \times \bbC^*$. This is the SW curve for the 5d theory compactified on a circle. We could eliminate three parameters by rescaling $F$, $\alpha$, $\beta$. The parameters $C_{i,j}$ are either Coulomb branch moduli or parameters like masses, couplings, etc\dots The number of moduli is given by the internal dots. The SW differential is defined by the holomorphic 2-form $d\lambda_{SW} = \Omega = d\log\alpha \wedge d\log\beta$.

Once the 5d theory is compactified on a circle, we obtain the 4d theory at low energies. The 4d limit arises as $L_B \to 0$, which means that the circle $L_A$ decompactifies. On the other hand the (classical) 4d coupling is
\be
g_{4d}^2 = \frac{g_{sB} \alpha'}{L_B L_w} = \frac{L_t}{(2\pi)^2L_w} \;,
\ee
where $L_w$ is the characteristic size of the web. To keep the coupling fixed, we take the web of the same
size as the M-theory circle. However the field theory is essentially independent of $l_p$, and we can send it to zero. In this way, the non-perturbative 4d dynamics is captured by weakly coupled M-theory \cite{Witten:1997ep}. Summarizing, the 4d limit corresponds to decompactifying one circle of the M-theory torus and scaling the parameters $C_{i,j}$ in such a way to keep the curve finite.

\medskip

To be concrete, the curve for the compactified 5d $N$-junction (see the toric diagram in figure \ref{fig: junctions}) is
\be \label{5d curve junction}
0 = F(\alpha,\beta) = \sum_{i,j \geq 0, \;\; i+j \leq N} C_{i,j} \, \alpha^i \beta^j \;.
\ee
To get the 4d limit, we first of all do the following redefinition:
\be
\alpha = t \, e^{\epsilon w} \qquad\qquad\qquad \beta = (t-1) \, e^{\epsilon w} \;,
\ee
in terms of which the SW differential is $d\lambda_{SW} = \epsilon \, t^{-1} (t-1)^{-1} \, dw \wedge dt$. To decompactify a circle, which will come from a combination of $\alpha$ and $\beta$, we will take $\epsilon \to 0$. The curve in terms of $t$ and $w$ reads:
\be
0 = F(w,t) = \sum_{i,j \geq 0, \;\; i+j \leq N} C_{i,j} \, e^{(i+j) \, \epsilon w} \; \sum_{k=0}^j (-1)^k \, \binom{k}{j} \,  t^{i+j-k} \;.
\ee
We can change indices to $l = i+j-k$ and $p=i+j$ to reorganize the summation in powers of $t$:
\be
0 = F(w,t) = \sum_{l=0}^N t^l \; \sum_{p=l}^N (-1)^{p-l} \, e^{p \, \epsilon w} \; \sum_{i=0}^l \binom{p-l}{p-i} \, C_{i,\, p-i} \;.
\ee
Now we take a scaling limit $\epsilon \to 0$ allowing the coefficients $C_{i,j}$ to diverge as $1/\epsilon$ at some power, as long as this does not lead to divergences in the curve.

Consider the coefficient of $t^N$ ($l=N$ in the first summation): it will be some power series in $\epsilon w$ whose coefficients are linear functions of the $C_{i,j}$. However $p=N$, therefore only one linear combination of the $C_{i,j}$ appears, multiplying the whole series expansion of $e^{N \epsilon w}$.
Such single combination must be finite in the $\epsilon \to 0$ limit: we get $a_{N,0}\, t^N$, where $a_{N,0} = \sum_{i=0}^N C_{i,N-i}$, without powers of $w$. Now consider the coefficient of $t^{N-1}$: this time there are two linear combinations of the $C_{i,j}$ appearing in front of two exponential functions of $w$, corresponding to $p=N,\, N-1$. We can set the coefficients in such a way that the two linear combinations diverge as $1/\epsilon$, so that the term $a_{N-1,1} \, t^{N-1} \, \epsilon w$ survives but the term $a_{N-1,0} \, t^{N-1}$ does not diverge. In general the coefficient of $t^l$ is the sum of $N-l$ linear combinations of the $C_{i,j}$ multiplying exponential functions of $\epsilon w$, for $p= N, \dots , l$. This allows to set the $C_{i,j}$ in such a way that all linear combinations diverge as $1/\epsilon^{N-l}$, but in the power series of the coefficient all divergences cancel. Taking $\epsilon \to 0$ we are left with the $N-l$ terms $(a_{l,N-l} \, w^{N-l} + \dots + a_{l,0})\, t^l$. Eventually, we get the most general polynomial in $t$ and $w$ with combined degree $N$:
\be
0 = F(t,w) = P_0 \, w^N + P_1(t) \, w^{N-1} + \dots + P_{N-1}(t) \, w + P_N(t) \;,
\ee
where $P_j$ are polynomials of degree $j$. Notice that the total number of parameters is $(N+1)(N+2)/2$, as in the 5d curve (\ref{5d curve junction}). We can then rescale $w$ to set $P_0=1$, and shift it $w \to w - P_1(t)/N$ to set $P_1(t)=0$. As remarked in \cite{Gaiotto:2009we}, keeping the SW differential fixed under such a shift corresponds to a harmless redefinition of the flavor currents.

Finally, we introduce a new coordinate $x=t^{-1}(t-1)^{-1}w$ in terms of which the curve reads:
\be
x^N = \frac{P_2(t)}{t^2(t-1)^2} \, x^{N-2} + \dots + \frac{P_{N-1}(t)}{t^{N-1}(t-1)^{N-1}} \, x + \frac{P_N(t)}{t^N(t-1)^N} \;.
\ee
The polynomials $P_j(t)$ encode all parameters, \ie{} Coulomb branch moduli and mass deformations, of the $T[A_{N-1}]$ theory. We can rewrite it in a more inspiring way as
\be
\label{curve 4d}
x^N = \phi_2 \, x^{N-2} + \dots + \phi_{N-1} \, x + \phi_N \;,
\ee
where
\be
\Phi_k \equiv \phi_k \, dt^k = \frac{P_k(t)}{t^k (t-1)^k} \, dt^k
\ee
are rank $k$ holomorphic differentials on the sphere with poles of order $k$ at $t=0,\, 1,\, \infty$. The holomorphic 2-form is $d\lambda_{SW} = dx \wedge dt$, and in fact
\be
\lambda_{SW} = x\, dt \;.
\ee
The curve (\ref{curve 4d}) and the differential agree with what found in \cite{Gaiotto:2009we}.

\section{$E_n$ theories}
\label{sec: exp theories}

Here we provide a brief review of what was known about
non-gravitational supersymmetric theories with $E_n$ flavor symmetry.
In the main part of our paper we provided new, dual realization of these theories
in four and five dimensions.
The paper \cite{Ganor:1996pc} would be a good starting point to the huge literature
on this subject.

The most basic theory is the six-dimensional $(1,0)$-theory with $E_8$ global symmetry,
which is realized on an M5-brane very close to the 9-brane `at the end of the world'
of the heterotic M-theory. It has one tensor multiplet, the scalar component of which
measures the distance between the M5-brane and the end of the world.
The M5-brane can be absorbed into the end of the world, becoming an $E_8$-instanton
which describes the Higgs branch of the theory; therefore this theory arises
on a point-like $E_8$-instanton of the $E_8\times E_8$ heterotic string.
This system has a dual geometric realization as a compactification of F-theory
with vanishing $S^2$ in the base. The total space contains the ninth del Pezzo.

Compactification of this theory on $S^1$ gives
five-dimensional theories with $E_n$ flavor symmetry.
On the side using branes,  we have a D4-brane probing a stack of an O8-plane and a few D8-branes
such that the dilaton diverges at the orientifold.
The Coulomb branch is real one dimensional,
and at the origin the Higgs branch emanates,
which describes the process where a D4-brane turns into an $E_{6,7,8}$-instanton.
On the purely geometric side, it is given by compactification of M-theory on
Calabi-Yau's containing vanishing $6,7,8$-th del Pezzo.
The Higgs branch is realized here by the extremal transition of the Calabi-Yau.

Further compactification on $S^1$ gives
four-dimensional $\cN=2$ SCFT's with $E_n$ flavor symmetry,
originally discussed in \cite{Minahan:1996fg,Minahan:1996cj}.
Let us discuss them using branes.  T-duality along the compactified $S^1$ gives us
a D3-brane probing a system of O7-planes and D7-branes.
It is then better to use the F-theory language which clearly describes the non-perturbative
properties of 7-branes, which we provide below in slightly more details.

We start from a flat 10-dimensional spacetime of type IIB or F-theory,
and put a 7-brane of type $E_n$
extending along $x^{0,1,2,3}$ and $x^{6,7,8,9}$ on which an 8d $E_n$ gauge theory lives.
We probe this background with $N$ D3-branes, extending along $x^{0,1,2,3}$.
The worldvolume theory on the D3-branes is the rank-$N$ $E_n$ theory.

Its Coulomb branch has  $N$ complex dimensions, parameterized by
the positions of the $N$ D3-branes along the directions $x^{4,5}$.
The scaling dimensions of these Coulomb branch operators are
\begin{equation}
\Delta,\, 2\Delta,\,\ldots,\, N\Delta
\label{tac:dimensions}
\end{equation}
where $\Delta$ is the dimension of the lowest dimension operator,
\begin{equation}
\Delta_{E_6}=3 \;,\qquad
\Delta_{E_7}=4 \;,\qquad
\Delta_{E_8}=6 \;.
\end{equation}

One way to understand this result is to recall that
the 7-brane is a codimension-two object and produces
a deficit angle.
The transverse space to a 7-brane with $E_{6,7,8}$ gauge group
is of the form $\bC/\bZ_{3,4,6}$.
The coordinate $z$ of $\bC$ parameterizes the Coulomb branch
and is of scaling dimension 1. The natural coordinate around the 7-brane is then
$u = z^{3,4,6}$ respectively, whose dimension is $3,4,6$.
When there are $N$ D3-branes, we have parameters $u_i$ for each D3-brane. However
the D3-branes are indistinguishable, therefore the Coulomb branch is parameterized
invariantly by symmetric polynomials of $u_i$, whose dimensions are exactly as shown in
\eqref{tac:dimensions}.

Central charges of these theories was found in  \cite{Cheung:1997id,Aharony:2007dj}.
In particular, the two-point function of the $E_n$ currents are characterized by the number
\begin{equation}
k_{E_n}= 2N \Delta
\end{equation}
which is normalized so that one hypermultiplet in the fundamental of $\SU(N)$
contributes 2 to $k_{\SU(N)}$.

When a D3-brane hits the 7-brane, the former can be absorbed into the latter
as an instanton. In the four-dimensional language, the Higgs branch emanates
from the origin of the Coulomb branch and it is identified with the $N$-instanton
moduli space of the gauge group $E_n$.  The center-of-mass of the instanton
configuration along $x^{6,7,8,9}$ is completely decoupled from
the rest of the system, so the true Higgs branch is the so-called `centered moduli space.'
The quaternionic dimension of this space is given by
\begin{equation}
Nh^\vee_{E_n}-1 \;,
\label{tac:higgsdim}
\end{equation}
where $h^\vee_{E_n}$ is the dual Coxeter number of the respective group,
given by
\begin{equation}
h^\vee_{E_6}=12 \;,\qquad
h^\vee_{E_7}=18 \;,\qquad
h^\vee_{E_8}=30 \;.
\end{equation}

In a similar manner we can consider rank-$N$ versions of
the five-dimensional $E_{6,7,8}$ theories and six-dimensional $E_8$ theory,
by putting $N$ D4-branes or $N$ M5-branes probing the O8-plane
or the `end-of-the world' brane, respectively.
These theories have real $N$-dimensional Coulomb branch,
and the Higgs branch is the $N$-instanton moduli space of $E_{6,7,8}$
which describes the process of branes being absorbed into branes as instantons.
Dual, purely geometric realizations of these higher-rank versions have not been well understood.

\end{document}